\pdfoutput=1
\newcommand*{\ATLASLATEXPATH}{./}
\documentclass[cernpreprint,texlive=2016,txfonts,texmf,UKenglish]{atlasdoc}
\makeatletter
\DeclareOldFontCommand{\rm}{\normalfont\rmfamily}{\mathrm}
\DeclareOldFontCommand{\sf}{\normalfont\sffamily}{\mathsf}
\DeclareOldFontCommand{\tt}{\normalfont\ttfamily}{\mathtt}
\DeclareOldFontCommand{\bf}{\normalfont\bfseries}{\mathbf}
\DeclareOldFontCommand{\it}{\normalfont\itshape}{\mathit}
\DeclareOldFontCommand{\sl}{\normalfont\slshape}{\@nomath\sl}
\DeclareOldFontCommand{\sc}{\normalfont\scshape}{\@nomath\sc}

\usepackage[biblatex=false]{\ATLASLATEXPATH atlaspackage}
\usepackage{\ATLASLATEXPATH atlasbiblatex}

\usepackage{\ATLASLATEXPATH atlascontribute}

\usepackage{\ATLASLATEXPATH atlasphysics}

\graphicspath{{./}{./}}

\AtlasTitle{Measurement of longitudinal flow decorrelations in Pb+Pb collisions at $\sqrt{s_{\textrm{NN}}}=2.76$ and 5.02 TeV with the ATLAS detector}
\AtlasAbstract{%
Measurements of longitudinal flow correlations are presented for charged particles in the pseudorapidity range $|\eta|<2.4$ using 7 $\mu$b$^{-1}$ and 470 $\mu$b$^{-1}$ of Pb+Pb collisions at $\sqrt{s_{\textrm{NN}}}=2.76$ and 5.02 TeV, respectively, recorded by the ATLAS detector at the LHC. It is found that the correlation between the harmonic flow coefficients $v_n$ measured in two separated $\eta$ intervals does not factorise into the product of single-particle coefficients, and this breaking of factorisation, or flow decorrelation, increases linearly with the $\eta$ separation between the intervals. The flow decorrelation is stronger at 2.76 TeV than at 5.02 TeV. Higher-order moments of the correlations are also measured, and the corresponding linear coefficients for the $k^{\textrm{th}}$-moment of the $v_n$ are found to be proportional to $k$ for $v_3$, but not for $v_2$. The decorrelation effect is separated into contributions from the magnitude of $v_n$ and the event-plane orientation, each as a function of $\eta$. These two contributions are found to be comparable. The longitudinal flow correlations are also measured between $v_n$ of different order in $n$. The decorrelations of $v_2$ and $v_3$ are found to be independent of each other, while the decorrelations of $v_4$ and $v_5$ are found to be driven by the nonlinear contribution from $v_2^2$ and $v_2v_3$, respectively.}

\author{The ATLAS Collaboration}

\PreprintIdNumber{CERN-PH-EP-2017-191}

\AtlasDate{\today}

\AtlasJournal{Eur. Phys. J. C}

\usepackage{float} 
\usepackage{amssymb}
\usepackage{url}
\usepackage{harpoon}
\usepackage{amsmath}
\usepackage{MnSymbol}
\usepackage{multirow}
\usepackage{bm}
\usepackage{cite}
\newcommand{\lr}[1]{\left\langle #1\right\rangle}

\newcommand{\etaref}{\eta_{\mathrm{ref}}}

\newcommand{\nchrec}{\mbox{$N_{\mathrm{ch}}^{\mathrm{rec}}$}}

\newcommand{\npart}{\mbox{$N_{\mathrm{part}}$}}

\newcommand{\sqrtsnn}{\mbox{$\sqrt{s_{\mathrm{NN}}}$}}

\hypersetup{pdftitle={ATLAS document},pdfauthor={The ATLAS Collaboration}}
\begin{document}

\maketitle



\section{Introduction}
\label{intro}

Heavy-ion collisions at RHIC and the LHC create hot, dense matter whose space-time evolution is well described by relativistic viscous hydrodynamics~\cite{Gale:2013da,Heinz:2013th}. Owing to strong event-by-event (EbyE) density fluctuations in the initial state, the space-time evolution of the produced matter also fluctuates event by event. These fluctuations lead to correlations of particle multiplicity in momentum space in both the transverse and longitudinal directions with respect to the collision axis. Studies of particle correlations in the transverse plane have revealed strong harmonic modulation of the particle densities in the azimuthal angle: d$N/{\textrm d}\phi\propto1+2\sum_{n=1}^{\infty}v_{n}\cos n(\phi-\Phi_{n})$, where $v_n$ and $\Phi_n$ represent the magnitude and event-plane angle of the $n^{\mathrm{th}}$-order harmonic flow. The measurements of harmonic flow coefficients $v_n$ and their EbyE fluctuations, as well as the correlations between $\Phi_{n}$ of different order~\cite{Adare:2011tg,ALICE:2011ab,Aad:2012bu,Chatrchyan:2013kba,Aad:2013xma,Aad:2014fla,Aad:2015lwa}, have placed important constraints on the properties of the dense matter and on transverse density fluctuations in the initial state~\cite{Luzum:2012wu,Teaney:2010vd,Gale:2012rq,Niemi:2012aj,Qiu:2012uy,Teaney:2013dta}.

Most previous flow studies assumed that the initial condition and space-time evolution of the matter are boost-invariant in the longitudinal direction. Recent model studies of two-particle correlations as a function of pseudorapidity $\eta$ revealed strong EbyE fluctuations of the flow magnitude and phase between two well-separated pseudorapidities, i.e.\ $v_n(\eta_1)\neq v_n(\eta_2)$  (forward-backward or FB asymmetry) and $\Phi_{n}(\eta_1)\neq \Phi_{n}(\eta_2)$ (event-plane twist)~\cite{Bozek:2010vz,Xiao:2012uw,Jia:2014ysa}. The CMS Collaboration proposed an observable based on the ratio of two correlations: the correlation between $\eta$ and $\etaref$ and the correlation between $-\eta$ and $\etaref$. This ratio is sensitive to the correlation between $\eta$ and $-\eta$~\cite{Khachatryan:2015oea}. The CMS results show that the longitudinal fluctuations lead to a linear decrease of the ratio with $\eta$, and the slope of the decrease shows a strong centrality dependence for elliptic flow $v_2$ but very weak dependences for $v_3$ and $v_4$. This paper extends the CMS result by measuring several new observables based on multi-particle correlations in two or more $\eta$ intervals~\cite{Jia:2017kdq}. These observables are sensitive to the EbyE fluctuations of the initial condition in the longitudinal direction. They are also sensitive to nonlinear mode-mixing effects, e.g.\ $v_4$ contains nonlinear contributions that are proportional to $v_2^2$~\cite{Gardim:2011xv,Teaney:2012ke,Aad:2014fla,Aad:2015lwa,Acharya:2017zfg}. Furthermore, the measurements are performed at two nucleon--nucleon centre-of-mass collision energies, $\sqrt{s_{\textrm{NN}}}=2.76$~TeV and 5.02~TeV, to evaluate the $\sqrt{s_{\textrm{NN}}}$ dependence of the longitudinal flow fluctuations. Recent model calculations predict an increase of longitudinal flow fluctuations at lower $\sqrt{s_{\textrm{NN}}}$~\cite{Pang:2015zrq}. Therefore, measurements of these observables at two collision energies can provide new insights into the initial condition along the longitudinal direction and should help in the development of full three-dimensional viscous hydrodynamic models.

Using these new observables, this paper improves the study of the longitudinal dynamics of collective flow in three ways. Firstly, the CMS measurement, which is effectively the first moment of the correlation between $v_n$ in separate $\eta$ intervals, is extended to the second and the third moments. Secondly, a correlation between four different $\eta$ intervals is measured to estimate the contributions from the fluctuations of $v_n$ amplitudes as well as the contributions from fluctuations of $\Phi_{n}$. Thirdly, correlations between harmonics of different order are also measured, e.g.\ between $v_2$ and $v_4$ in different $\eta$ intervals, to investigate how mode-mixing effects evolve with rapidity. In this way, this paper presents a measurement of flow decorrelation involving $v_2$, $v_3$, $v_4$ and $v_5$, using Pb+Pb collisions at $\sqrt{s_{\textrm{NN}}}=2.76$ and 5.02~TeV.

\section{Observables}
\label{sec:obs}

This section gives a brief summary of the observables measured in this paper, further details can be found in Refs.~\cite{Khachatryan:2015oea,Bozek:2015bna,Jia:2017kdq}. The azimuthal anisotropy of the particle production in an event is conveniently described by harmonic flow vectors $ {\bm V}_n = v_n e^{{\textrm i}n\Phi_n}$~\footnote{As in several previous analyses~\cite{Luzum:2013yya,Bilandzic:2010jr}, a complex number is used to represent the real two-dimensional flow vector.}, where $v_n$ and $\Phi_n$ are the magnitude and phase (or event plane), respectively. The $ {\bm V}_n$ are estimated from the observed per-particle normalised flow vector ${\bm q}_n$~\cite{Aad:2012bu}:
\begin{eqnarray}
\label{eq:1}
{\bm q}_{n} \equiv \frac{\sum_i  w_ie^{{\textrm i}n\phi_i}}{\sum_iw_i}\;.
\end{eqnarray}
The sums run over all particles in a given $\eta$ interval of the event, and $\phi_i$ and $w_i$ are the azimuthal angle and the weight assigned to the $i^{\textrm{th}}$ particle, respectively. The weight accounts for detector non-uniformity and tracking inefficiency. 

\begin{figure}[h!]
\begin{center}
\includegraphics[width=1\linewidth]{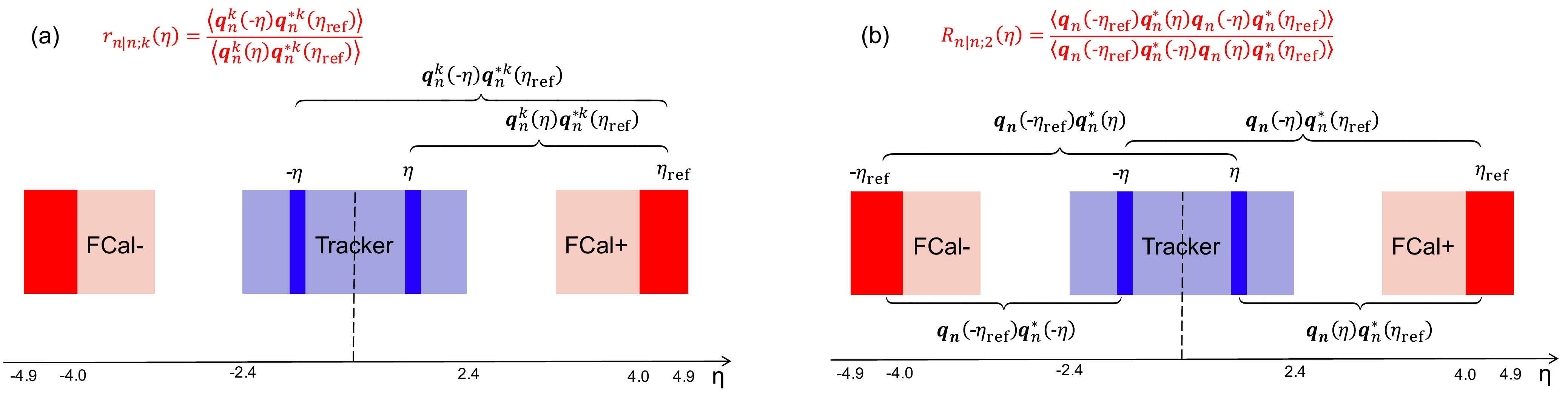}
\end{center}\vspace*{-0.4cm}
\caption{\label{fig:0} Schematic illustration of the procedure for constructing the corrlators $r_{n|n;k}(\eta)$ Eq.~\eqref{eq:2} (left panel) and $R_{n|n;2}(\eta)$ Eq.~\eqref{eq:6} (right panel). The acceptance coverages for the ATLAS tracker used for $\eta$ and reference detector used for $\etaref$ are discussed in Section~\ref{sec:corr}.}
\end{figure}

The longitudinal flow fluctuations are studied using the correlation between the $k^{\textrm {th}}$-moment of the $n^{\textrm {th}}$-order flow vectors in two different $\eta$ intervals, averaged over events in a given centrality interval, $r_{n|n;k}$, for $k=1$,2,3:
\begin{eqnarray}
\label{eq:2}
r_{n|n;k}(\eta) = \frac{\lr{{\bm q}_n^k (-\eta) {\bm q}_n^{*k}(\etaref)}}{\lr{{\bm q}_n^k (\eta){\bm q}_n^{*k}(\etaref)}} = \frac{\lr{\left[v_n(-\eta) v_n(\etaref)\right]^k \cos kn(\Phi_n(-\eta)-\Phi_n(\etaref))}}{\lr{\left[v_n(\eta) v_n(\etaref)\right]^k \cos kn(\Phi_n(\eta)-\Phi_n(\etaref))}}\;,
\end{eqnarray}
where $\etaref$ is the reference pseudorapidity common to the numerator and the denominator, the subscript ``$n|n;k$'' denotes the $k^{\textrm{th}}$-moment of the flow vectors of order $n$ at $\eta$, combined with the $k^{\textrm{th}}$ moment of the conjugate of the flow vector of order $n$ at $\etaref$. The sine terms vanish in the last expression in Eq.~\eqref{eq:2} because any observable must be an even function of $\Phi_n(-\eta)-\Phi_n(\etaref)$. A schematic illustration of the choice of the $\eta$ ($|\eta|<2.4$) and $\etaref$ ($4.0<|\etaref|<4.9$) to be discussed in Section~\ref{sec:corr}, as well as the relations between different flow vectors, are shown in the left panel of Figure~\ref{fig:0}. This observable is effectively a 2k-particle correlator between two subevents as defined in Ref.~\cite{Jia:2017hbm}, and the particle multiplets containing duplicated particle indices are removed using the cumulant framework, with particle weights taken into account~\cite{Jia:2017kdq}.

 The observable measured by the CMS Collaboration~\cite{Khachatryan:2015oea} corresponds to $k=1$, i.e.\ $r_{n|n;1}$ . It should be noted that $\lr{{\bm q}_n}=0$ because the event plane changes randomly from event to event. Hence a direct study of the correlation between $+\eta$ and $-\eta$ via a quantity such as $\lr{{\bm q}_n (+\eta) {\bm q}_n^{*}(-\eta)}/(\lr{{\bm q}_n (+\eta)}\lr{{\bm q}_n^{*} (-\eta)})$ is not possible. One could also consider a quantity like $\lr{{\bm q}_n (+\eta) {\bm q}_n^{*}(-\eta)}/\left(\lr{q_n^2 (\eta)}\lr{q_n^2 (-\eta)}\right)^{1/2}$, but the denominator would be affected by short-range correlations. Hence, it is preferable to work with quantities of the type used in Eq.~\eqref{eq:2}, which give a correlator sensitive to the flow decorrelation between $\eta$ and $-\eta$ through the reference flow vector ${\bm q}_n^{k}(\etaref)$.

One important feature of Eq.~\eqref{eq:2} is that the detector effects at $\etaref$ are expected to cancel out to a great extent (see Section~\ref{sec:corr}). To ensure a sizeable pseudorapidity gap between the flow vectors in both the numerator and denominator of Eq.~\eqref{eq:2}, $\etaref$ is usually chosen to be at large pseudorapidity, e.g.\ $\etaref>4$ or $\etaref<-4$, while the pseudorapidity of ${\bm q}_n(-\eta)$ and ${\bm q}_n(\eta)$ is usually chosen to be close to mid-rapidity, $|\eta|<2.4$. If flow harmonics from multi-particle correlations factorise into single-particle flow harmonics, e.g. \mbox{$\lr{{\bm V}_n^k (\eta) {\bm V}_n^{*k}(\etaref)}^2=\lr{ v_n^{2k} (\eta)} \lr{v_n^{2k}(\etaref)}$}, then it is expected that $r_{n|n;k}(\eta)=1$. Therefore, a value of $r_{n|n;k}(\eta)$ different from 1 implies a factorisation-breaking effect due to longitudinal flow fluctuations, and such an effect is generally referred to as ``flow decorrelation''.

Based on the CMS measurement~\cite{Khachatryan:2015oea} and arguments in Ref.~\cite{Jia:2017kdq}, the observable $r_{n|n;k}(\eta)$ is expected to be approximately a linear function of $\eta$ with a negative slope, and is sensitive to both the asymmetry in the magnitude of $v_n$ and the twist of the event-plane angles between $\eta$ and $-\eta$: 
\begin{eqnarray}
\label{eq:3}
r_{n|n;k}(\eta) \approx 1-2F_{n;k}^{\textrm r}\eta,\;\; F_{n;k}^{\textrm r}= F_{n;k}^{\textrm{asy}}+F_{n;k}^{\textrm{twi}}\;,
\end{eqnarray}
where $F_{n;k}^{\textrm{asy}}$ and $F_{n;k}^{\textrm{twi}}$ represent the contribution from FB $v_n$ asymmetry and event-plane twist, respectively. The $r_{n|n;k}$ results obtained in Ref.~\cite{Khachatryan:2015oea} were for $k=1$ and $n=2$, 3, 4. The measured $F_{n;1}^{\textrm r}$ show only a weak dependence on $\etaref$ for $\etaref>3$ or $\etaref<-3$ at the LHC. Measuring $r_{n|n;k}$ for $k>1$ provides new information on how the $v_n$ asymmetry and event-plane twist fluctuate event by event. 

If the amount of decorrelation for the $k^{\textrm{th}}$-moment of the flow vector is proportional to $k$, it can be shown that~\cite{Jia:2017kdq}:
\begin{eqnarray}
\label{eq:5}
r_{n|n;k}\approx r_{n|n;1}^k\;,\;\;F_{n;k}^{\textrm r}\approx k F_{n;1}^{\textrm r}\;.
\end{eqnarray}
Deviations from Eq.~\eqref{eq:5} are sensitive to the detailed EbyE structure of the flow fluctuations in the longitudinal direction.

To estimate the separate contributions of the asymmetry and twist effects, a new observable involving correlations of flow vectors in four $\eta$ intervals is used~\cite{Jia:2017kdq}:
\begin{eqnarray}
\label{eq:6}
\nonumber
R_{n|n;2}(\eta) &=& \frac{\lr{{\bm q}_n(-\etaref){\bm q}_n^{*} (\eta) {\bm q}_n (-\eta) {\bm q}_n^*(\etaref)}}{\lr{{\bm q}_n(-\etaref){\bm q}_n^* (-\eta){\bm q}_n (\eta) {\bm q}_n^*(\etaref)}}\\\nonumber
&=&\frac{\lr{v_n(-\etaref)v_n(-\eta)v_n(\eta)v_n(\etaref)\cos n\left[\Phi_n(-\etaref)-\Phi_n(\etaref)+(\Phi_n(-\eta)-\Phi_n(\eta))\right]}}{\lr{v_n(-\etaref)v_n(-\eta)v_n(\eta)v_n(\etaref)\cos n\left[\Phi_n(-\etaref)-\Phi_n(\etaref)-(\Phi_n(-\eta)-\Phi_n(\eta))\right]}}\;,\\
\end{eqnarray}
where the notation ``2'' in the subscript indicates that there are two ${\bm q}_n$ and two ${\bm q}_n^{*}$ in the numerator and denominator. A schematic illustration of the relations between different flow vectors is shown in the right panel of Figure~\ref{fig:0}. Since the effect of an asymmetry is the same in both the numerator and the denominator, this correlator is mainly sensitive to the event-plane twist effects:
\begin{eqnarray}
\label{eq:7}
R_{n|n;2} (\eta) \approx 1-2F_{n;2}^{\textrm{R}}\eta\;, F_{n;2}^{\textrm{R}}= F_{n;2}^{\textrm{twi}}\;.
\end{eqnarray}
Therefore, the asymmetry and twist contributions can be estimated by combining Eqs.~\eqref{eq:3} and \eqref{eq:7}.

Measurements of longitudinal flow fluctuations can also be extended to correlations between harmonics of different order:
\begin{eqnarray}
\label{eq:8}
r_{2,3|2,3}(\eta) &=& \frac{\lr{{\bm q}_2 (-\eta){\bm q}_2^{*}(\etaref){\bm q}_3 (-\eta){\bm q}_3^{*}(\etaref)}}{\lr{{\bm q}_2 (\eta){\bm q}_2^{*}(\etaref){\bm q}_3 (\eta){\bm q}_3^{*}(\etaref)}}\;,\\\label{eq:9}
r_{2,2|4}(\eta) &=& \frac{\lr{{\bm q}_2^2 (-\eta) {\bm q}_4^{*}(\etaref)}+\lr{{\bm q}_2^2 (\etaref) {\bm q}_4^{*}(-\eta)}}{\lr{{\bm q}_2^2 (\eta){\bm q}_4^{*}(\etaref)}+\lr{{\bm q}_2^2 (\etaref){\bm q}_4^{*}(\eta)}}\;,\\\label{eq:10}
r_{2,3|5}(\eta) &=& \frac{\lr{{\bm q}_2(-\eta){\bm q}_3(-\eta) {\bm q}_5^{*}(\etaref) }+\lr{{\bm q}_2(\etaref){\bm q}_3(\etaref) {\bm q}_5^{*}(-\eta) }}{\lr{{\bm q}_2 (\eta){\bm q}_3 (\eta){\bm q}_5^{*}(\etaref)}+\lr{{\bm q}_2 (\etaref){\bm q}_3 (\etaref){\bm q}_5^{*}(\eta)}}\;,
\end{eqnarray}
where the comma in the subscripts denotes the combination of ${\bm q}_n$ of different order. If the longitudinal fluctuations for ${\bm V}_2$ and ${\bm V}_3$ are independent of each other, one would expect \mbox{$r_{2,3|2,3}= r_{2|2;1}r_{3|3;1}$}~\cite{Jia:2017kdq}. On the other hand, $r_{2,2|4}$ and $r_{2,3|5}$ are sensitive to the $\eta$ dependence of the correlations between $v_n$ and event planes of different order, for example \mbox{$\lr{{\bm q}_2^2 (-\eta) {\bm q}_4^{*}(\etaref)} = \lr{v_2^2(-\eta) v_4(\etaref) \cos 4(\Phi_2(-\eta)-\Phi_4(\etaref))}$}. Correlations between different orders have been measured previously at the LHC~\cite{Aad:2014fla,Aad:2015lwa,ALICE:2016kpq,Acharya:2017zfg}.

It is well established that the ${\bm V}_4$ and ${\bm V}_5$ in Pb+Pb collisions contain a linear contribution associated with initial geometry and mode-mixing contributions from lower-order harmonics due to nonlinear hydrodynamic response~\cite{Gardim:2011xv,Teaney:2012ke,Aad:2014fla,Aad:2015lwa,Qiu:2012uy}:
\begin{eqnarray}
\label{eq:11}
{\bm V}_4 = {\bm V}_{4\textrm{L}} +\chi_{4} {\bm V}_2^2\;,\;\;{\bm V}_5 = {\bm V}_{5\textrm{L}}+\chi_{5} {\bm V}_2{\bm V}_3\;,
\end{eqnarray}
where the linear component ${\bm V}_{n\textrm{L}}$ is driven by the corresponding eccentricity in the initial geometry~\cite{Teaney:2010vd}. If the linear component of $v_4$ and $v_5$ is uncorrelated with lower-order harmonics, i.e.\ ${\bm V}_2^2{\bm V}_{4\textrm{L}}^{*}\sim0$ and ${\bm V}_2{\bm V}_3{\bm V}_{5\textrm{L}}^{*}\sim0$, one expects~\cite{Jia:2017kdq}:
\begin{eqnarray}
\label{eq:12}
r_{2,2|4}\approx r_{2|2;2}\;,\;\;r_{2,3|5}\approx r_{2,3|2,3}\;.
\end{eqnarray}
Furthermore, using Eq.~\eqref{eq:11} the $r_{n|n;1}$ correlators involving $v_4$ and $v_5$ can be approximated by:
\begin{eqnarray}
\label{eq:13}
r_{4|4;1}(\eta) &\approx& \frac{\lr{{\bm V}_{4\textrm{L}}(-\eta){\bm V}_{4\textrm{L}}^{*}(\etaref)}+\chi_{4}^2\lr{{\bm V}_2^2 (-\eta) {\bm V}_2^{*2}(\etaref)}}{\lr{{\bm V}_{4\textrm{L}}(\eta){\bm V}_{4\textrm{L}}^{*}(\etaref)}+\chi_{4}^2\lr{{\bm V}_2^2 (\eta) {\bm V}_2^{*2}(\etaref)}}\;,\\\label{eq:12b}
r_{5|5;1}(\eta) &\approx& \frac{\lr{{\bm V}_{5\textrm{L}}(-\eta){\bm V}_{5\textrm{L}}^{*}(\etaref)}+\chi_{5}^2\lr{{\bm V}_2 (-\eta){\bm V}_2^{*}(\etaref){\bm V}_3 (-\eta){\bm V}_3^{*}(\etaref)}}{\lr{{\bm V}_{5\textrm{L}}(\eta){\bm V}_{5\textrm{L}}^{*}(\etaref)}+\chi_{5}^2\lr{{\bm V}_2 (\eta){\bm V}_2^{*}(\etaref){\bm V}_3 (\eta){\bm V}_3^{*}(\etaref)}}\;.\\\nonumber
\end{eqnarray}
Therefore, both the linear and nonlinear components are important for $r_{4|4;1}$ and $r_{5|5;1}$.

\section{ATLAS detector and trigger}
\label{sec:det}
The ATLAS detector~\cite{Aad:2008zzm} provides nearly full solid-angle coverage of the collision point with tracking detectors, calorimeters, and muon chambers, and is well suited for measurements of multi-particle correlations over a large pseudorapidity range.\footnote{ATLAS uses a right-handed coordinate system with its origin at the nominal interaction point (IP) in the centre of the detector and the $z$-axis along the beam pipe. The $x$-axis points from the IP to the centre of the LHC ring, and the $y$-axis points upward. Cylindrical coordinates $(r,\phi)$ are used in the transverse plane, $\phi$ being the azimuthal angle around the beam pipe. The pseudorapidity is defined in terms of the polar angle $\theta$ as $\eta=-\ln\tan(\theta/2)$.} The measurements were performed using the inner detector~(ID), minimum-bias trigger scintillators (MBTS), the forward calorimeters~(FCal), and the zero-degree calorimeters (ZDC). The ID detects charged particles within $|\eta| < 2.5$ using a combination of silicon pixel detectors, silicon microstrip detectors~(SCT), and a straw-tube transition-radiation tracker (TRT), all immersed in a 2 T axial magnetic field~\cite{Aad:ID}. An additional pixel layer, the ``insertable B-layer''~(IBL)~\cite{ATLAS-TDR-19} installed during the 2013-2015 shutdown between Run 1 and Run 2, is used in the 5.02~$\TeV$ measurements. The MBTS system detects charged particles over $2.1\lesssim|\eta|\lesssim3.9$ using two hodoscopes of counters positioned at $z = \pm$3.6 m. The FCal consists of three sampling layers, longitudinal in shower depth, and covers $3.2<|\eta|< 4.9$. The ZDC are positioned at $\pm$140~m from the IP, detecting neutrons and photons with $|\eta|>8.3$. 

This analysis uses approximately 7~$\mu\textrm{b}^{-1}$ and 470~$\mu\textrm{b}^{-1}$ of Pb+Pb data at $\sqrtsnn = 2.76$ and 5.02~$\TeV$, respectively, recorded by the ATLAS experiment at the LHC. The 2.76~TeV data were collected in 2010, while the 5.02~TeV data were collected in 2015. 

The ATLAS trigger system~\cite{Aad:2012xs} consists of a level-1 (L1) trigger implemented using a combination of dedicated electronics and programmable logic, and a high-level trigger (HLT) implemented in general-purpose processors. The trigger requires signals in both ZDC or either of the two MBTS counters. The ZDC trigger thresholds on each side are set below the maximum corresponding to a single neutron. A timing requirement based on signals from each side of the MBTS was imposed to remove beam backgrounds. This trigger selected 7~$\mu\textrm{b}^{-1}$ and 22~$\mu\textrm{b}^{-1}$ of minimum-bias Pb+Pb data at $\sqrtsnn = 2.76$~$\TeV$ and $\sqrtsnn = 5.02$~$\TeV$, respectively. To increase the number of recorded events from very central Pb+Pb collisions, a dedicated L1 trigger was used in 2015 to select events requiring the total transverse energy ($\Sigma \eT$) in the FCal to be more than 4.54~TeV. This ultra-central trigger sampled 470~$\mu$b$^{-1}$ of Pb+Pb collisions at 5.02~TeV and was fully efficient for collisions with centrality 0--0.1\% (see Section~\ref{sec:sel}).

\section{Event and track selection}
\label{sec:sel}
The offline event selection requires a reconstructed vertex with its $z$ position satisfying $|\zvtx|< 100$~mm. For the $\sqrtsnn = 2.76$~$\TeV$  Pb+Pb data, the selection also requires a time difference $|\Delta t| < 3$ ns between signals in the MBTS trigger counters on either side of the nominal centre of ATLAS to suppress non-collision backgrounds. A coincidence between the ZDC signals at forward and backward pseudorapidity is required to reject a variety of background processes such as elastic collisions and non-collision backgrounds, while maintaining high efficiency for inelastic processes. The fraction of events containing more than one inelastic interaction (pile-up) is estimated to be less than 0.1\% at both collision energies. The pile-up contribution is studied by exploiting the correlation between the transverse energy $\Sigma \eT$ measured in the FCal or the number of neutrons $N_n$ in the ZDC and the number of tracks associated with a primary vertex $\nchrec$. Since the distribution of $\Sigma \eT$ or $N_n$ in events with pile-up is broader than that for the events without pile-up, pile-up events are suppressed by rejecting events with an abnormally large $\Sigma \eT$ or $N_n$ as a function of $\nchrec$.

The event centrality~\cite{Aad:2011yr} is characterised by the $\Sigma \eT$ deposited in the FCal over the pseudorapidity range $3.2 < |\eta| < 4.9$ using a calibration employing the electromagnetic calorimeters to set the energy scale~\cite{Aad:2011he}. The FCal $\Sigma \eT$ distribution is divided into a set of centrality intervals. A centrality interval refers to a percentile range, starting at 0\% relative to the most central collisions. Thus the 0--5\% centrality interval, for example, corresponds to the most central 5\% of the events. The ultra-central trigger mentioned in Section~\ref{sec:det} selects events in the 0--0.1\% centrality interval with full efficiency. A Monte Carlo Glauber analysis~\cite{Miller:2007ri,Aad:2011yr} is used to estimate the average number of participating nucleons, $\npart$, for each centrality interval. The systematic uncertainty in $\npart$ is less than 1\% for centrality intervals in the range 0--20\%  and increases to 6\% for centrality intervals in the range 70--80\%. The Glauber model also provides a correspondence between the $\Sigma \eT$ distribution and sampling fraction of the total inelastic Pb+Pb cross section, allowing centrality percentiles to be set. For this analysis, a selection of collisions corresponding to 0--70\% centrality is used to avoid diffraction or other processes that contribute to very peripheral collisions. Following the convention used in heavy-ion analyses, the centrality dependence of the results in this paper is presented as a function of $\npart$.

Charged-particle tracks and primary vertices~\cite{Aad:2015wga} are reconstructed from hits in the ID. Tracks are required to have $\pT>0.5$~GeV and $|\eta|<2.4$. For the 2.76~TeV data, tracks are required to have at least nine hits in the silicon detectors with no missing pixel hits and not more than one missing SCT hit, taking into account the presence of known dead modules. For the 5.02~TeV data, tracks are required to have at least two pixel hits, with the additional requirement of a hit in the first pixel layer when one is expected, at least eight SCT hits, and at most one missing hit in the SCT. In addition, for both datasets, the point of closest approach of the track is required to be within 1~mm of the primary vertex in both the transverse and longitudinal directions~\cite{Aad:2011yk}.

The efficiency, $\epsilon(\pT,\eta)$, of the track reconstruction and track selection criteria is evaluated using Pb+Pb Monte Carlo events produced with the HIJING event generator~\cite{Gyulassy:1994ew}. The generated particles in each event were rotated in azimuthal angle according to the procedure described in Ref.~\cite{Masera:2009zz} to produce harmonic flow consistent with previous ATLAS measurements~\cite{ATLAS:2011ah,Aad:2012bu}. The response of the detector was simulated using $\GEANT$4~\cite{Agostinelli:2002hh,Aad:2010ah} and the resulting events are reconstructed with the same algorithms applied to the data. For the 5.02~TeV Pb+Pb data, the efficiency ranges from 75\% at $\eta\approx 0$ to about 50\% for $|\eta| > 2$ for charged particles with $\pT > 0.8$ GeV, falling by about 5\% as $\pT$ is reduced to 0.5 GeV. The efficiency varies more strongly with $\eta$ and event multiplicity. For $\pT > 0.8$~GeV, it ranges from 75\% at $\eta \approx 0$ to 50\% for $|\eta| > 2$ in peripheral collisions, while it ranges from 71\% at $\eta \approx 0$ to about 40\% for $|\eta| > 2$ in central collisions.  The tracking efficiency for the 2.76~TeV data has a similar dependence on $\pT$ and $\eta$. The efficiency is used in the particle weight, as described in Section~\ref{sec:corr}. However, because the observables studied are ratios (see Section~\ref{sec:obs}), uncertainties in detector and reconstruction efficiencies largely cancel. The rate of falsely reconstructed tracks (``fakes'') is also estimated and found to be significant only at $\pT<1$ GeV in central collisions, where its percentage per-track ranges from 2\% at $|\eta|<1$ to 8\% at the larger $|\eta|$. The fake rate drops rapidly for higher $\pT$ and towards more peripheral collisions. The fake rate is accounted for in the tracking efficiency correction following the procedure in Ref.~\cite{Aad:2014vba}.

\section{Data analysis}
\label{sec:corr}
Measurement of the longitudinal flow dynamics requires the calculation of the flow vector ${\bm q}_n$ via Eq.~\eqref{eq:1} in the ID and the FCal. The flow vector from the FCal serves as the reference ${\bm q}_n(\etaref)$, while the ID provides the flow vector as a function of pseudorapidity ${\bm q}_n(\eta)$.

In order to account for detector inefficiencies and non-uniformity, a particle weight for the $i^{\textrm{th}}$-particle in the ID for the flow vector from Eq.~\eqref{eq:1} is defined as:
\begin{eqnarray}\label{eq:b1}
w_i^{\textrm{ID}}(\eta,\phi,\pT) = d_{\textrm{ID}} (\eta,\phi)/\epsilon(\eta,\pT)\;,
\end{eqnarray}
similar to the procedure in Ref.~\cite{Aad:2014vba}. The determination of track efficiency $\epsilon(\eta,\pT)$ is described in Section~\ref{sec:sel}. The additional weight factor $d_{\textrm{ID}}(\eta,\phi)$ corrects for variation of tracking efficiency or non-uniformity of detector acceptance as a function of $\eta$ and $\phi$. For a given $\eta$ interval of 0.1, the distribution in azimuthal bins, $N(\phi,\eta)$, is built up from reconstructed charged particles summed over all events. The weight factor is then obtained as $d_{\textrm{ID}}(\eta,\phi) \equiv \lr{N(\eta)}/N(\phi,\eta)$, where $\lr{N(\eta)}$ is the average of $N(\phi,\eta)$. This ``flattening'' procedure removes most $\phi$-dependent non-uniformity from track reconstruction, which is important for any azimuthal correlation analysis. Similarly, the weight in the FCal for the flow vector from Eq.~\eqref{eq:1} is defined as:
\begin{eqnarray}\label{eq:b2}
w_i^{\textrm{FCal}}(\eta,\phi) = d_{\textrm{FCal}}(\eta,\phi)  E_{{\textrm{T}},i}\;,
\end{eqnarray}
where $E_{{\textrm{T}},i}$ is the transverse energy measured in the $\textrm{i}^{\textrm{th}}$ tower in the FCal at $\eta$ and $\phi$. The azimuthal weight $d_{\textrm{FCal}}(\eta,\phi)$ is calculated in narrow $\eta$ intervals in a similar way to what is done for the ID.  It ensures that the $E_{\textrm{T}}$-weighted distribution, averaged over all events in a given centrality interval, is uniform in $\phi$. The flow vectors ${\bm q}_n(\eta)$ and ${\bm q}_n(\etaref)$ are further corrected by an event-averaged offset: ${\bm q}_n -\lr{{\bm q}_n}_{\textrm{evts}}$~\cite{Aad:2014fla}. 

The flow vectors obtained after these reweighting and offset procedures are used in the correlation analysis. The correlation quantities used in $r_{n|n;k}$ are calculated as:
\begin{eqnarray}\label{eq:b3}
\lr{{\bm q}_n^k (\eta) {\bm q}_n^{*k}(\etaref)}\equiv \lr{{\bm q}_n^k (\eta) {\bm q}_n^{*k}(\etaref)}_{\textrm s}-\lr{{\bm q}_n^k (\eta) {\bm q}_n^{*k}(\etaref)}_{\textrm b}\;,
\end{eqnarray}
where subscripts ``s'' and ``b'' represent the correlator constructed from the same event (``signal'') and from the mixed-event (``background''), respectively. The mixed-event quantity is constructed by combining ${\bm q}_n^k (\eta)$ from each event with ${\bm q}_n^{*k}(\etaref)$ obtained in other events with similar centrality (within 1\%) and similar $\zvtx$ ($|\Delta\zvtx|<5$~mm). The $\lr{{\bm q}_n^k (\eta) {\bm q}_n^{*k}(\etaref)}_{\textrm b}$, which is typically more than two orders of magnitude smaller than the corresponding signal term, is subtracted to account for any residual detector non-uniformity effects that result from a correlation between different $\eta$ ranges. 

For correlators involving flow vectors in two different $\eta$ ranges, mixed events are constructed from two different events. For example, the correlation for $r_{2,3|5}$ is calculated as:
\begin{eqnarray}\label{eq:b4}
\lr{{\bm q}_2 (\eta) {\bm q}_3 (\eta) {\bm q}_5^{*}(\etaref)}\equiv \lr{{\bm q}_2 (\eta) {\bm q}_3 (\eta) {\bm q}_5^{*}(\etaref)}_{\textrm s}-\lr{{\bm q}_2 (\eta) {\bm q}_3 (\eta) {\bm q}_5^{*}(\etaref)}_{\textrm b}\;.
\end{eqnarray}
The mixed-event correlator is constructed by combining ${\bm q}_2 (\eta) {\bm q}_3 (\eta)$ from one event with ${\bm q}_5^{*}(\etaref)$ obtained in another event with similar centrality (within 1\%) and similar $\zvtx$ ($|\Delta\zvtx|<5$~mm). On the other hand, for correlators involving more than two different $\eta$ ranges, mixed events are constructed from more than two different events, one for each unique $\eta$ range. One such example is $R_{n|n;2}$, for which each mixed event is constructed from four different events with similar centrality and $\zvtx$.

Most correlators can be symmetrised. For example, in a symmetric system such as Pb+Pb collisions, the condition $\lr{{\bm q}_n^k (-\eta) {\bm q}_n^{*k}(\etaref)} = \lr{{\bm q}_n^k (\eta) {\bm q}_n^{*k}(-\etaref)}$ holds. So instead of Eq.~\eqref{eq:2}, the actual measured observable is: 
\begin{eqnarray}
\label{eq:b5}
r_{n|n;k}(\eta) = \frac{\lr{{\bm q}_n^k (-\eta) {\bm q}_n^{*k}(\etaref)+{\bm q}_n^k (\eta) {\bm q}_n^{*k}(-\etaref)}}{\lr{{\bm q}_n^k (\eta){\bm q}_n^{*k}(\etaref)+{\bm q}_n^k (-\eta){\bm q}_n^{*k}(-\etaref)}}\;.
\end{eqnarray}
The symmetrisation procedure also allows further cancellation of possible differences between $\eta$ and $-\eta$ in the tracking efficiency or detector acceptance.

Table~\ref{tab:obs} gives a summary of the set of correlators measured in this analysis. The analysis is performed in intervals of centrality and the results are presented as a function of $\eta$ for $|\eta|<2.4$. The main results are obtained using 5.02~TeV Pb+Pb data. The 2010 2.76~TeV Pb+Pb data are statistically limited, and are used only to obtain $r_{n|n;1}$ and $R_{n|n;2}$ to compare with results obtained from the 5.02~TeV data and study the dependence on collision energy. 
\begin{table}[t]
\centering
\caption{\label{tab:obs} The list of observables measured in this analysis.}
\begin{tabular}{|l|c|}\hline 
\multicolumn{1}{|c|}{Observables}& Pb+Pb datasets\tabularnewline\hline
$r_{n|n;k}$ \;\;\;for $n=2$,3,4 and $k=1$ & 2.76 and 5.02~TeV\tabularnewline\hline
$R_{n|n;2}$ \;\;for $n=2$,3        & 2.76 and 5.02~TeV\tabularnewline\hline
$r_{n|n;k}$ \;\;\;for $n=5$ and $k=1$ &  5.02~TeV\tabularnewline\hline
$r_{n|n;k}$ \;\;\;for $n=2$,3 and  $k=2$,3 & 5.02~TeV \tabularnewline\hline
$R_{n|n;2}$ \;\;for $n=4$          & 5.02~TeV\tabularnewline\hline
$r_{2,2|4}$,\;$r_{2,3|5}$,\;$r_{2,3|2,3}$  & 5.02~TeV\tabularnewline\hline
\end{tabular}
\end{table}

\begin{figure}[h!]
\begin{center}
\includegraphics[width=1\linewidth]{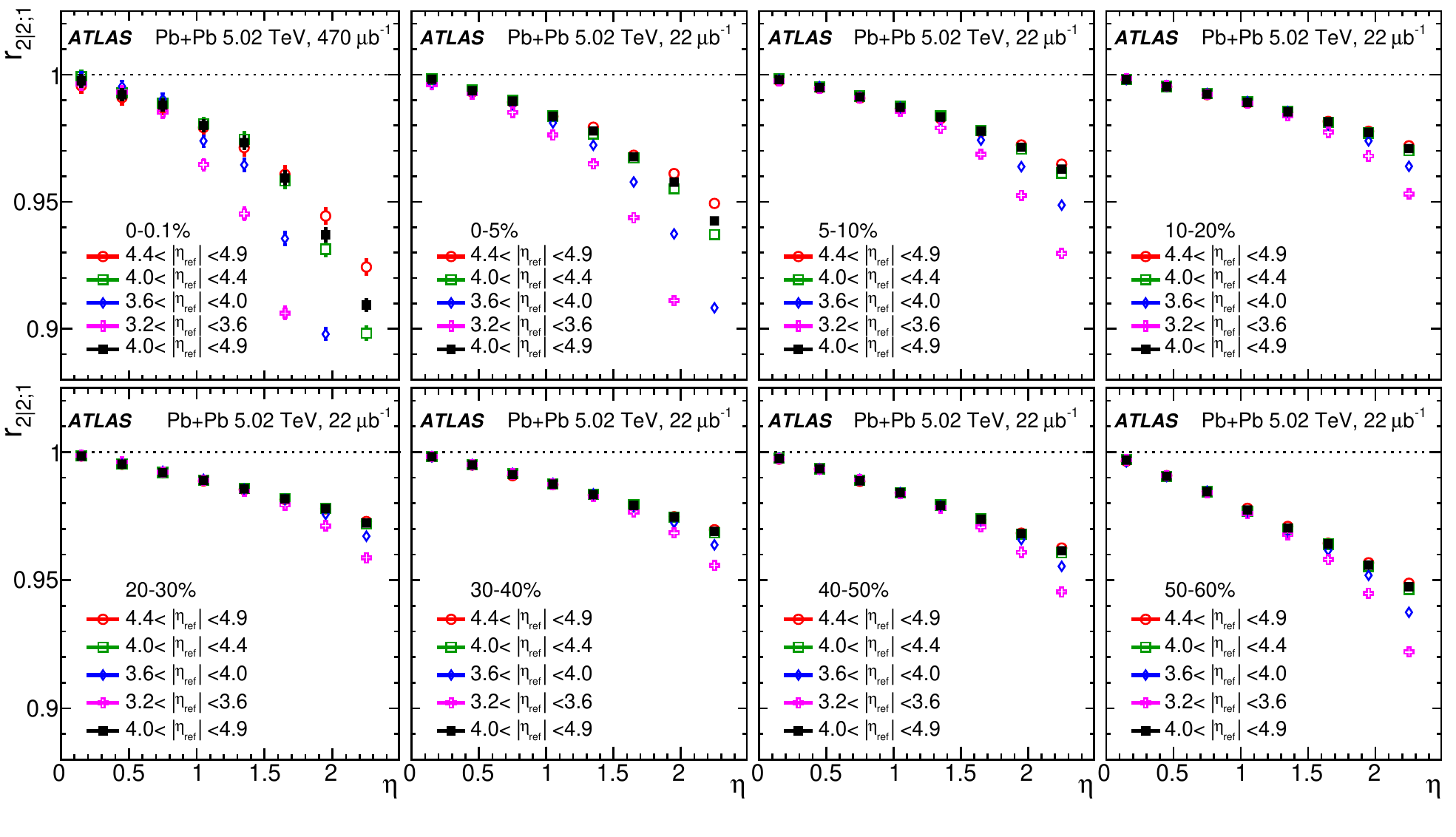}
\end{center}\vspace*{-0.4cm}
\caption{\label{fig:1} The $r_{2|2;1}(\eta)$ measured for several $\etaref$ ranges. Each panel shows the results for one centrality range. The error bars are statistical only.}
\end{figure}

\begin{figure}[h!]
\begin{center}
\includegraphics[width=1\linewidth]{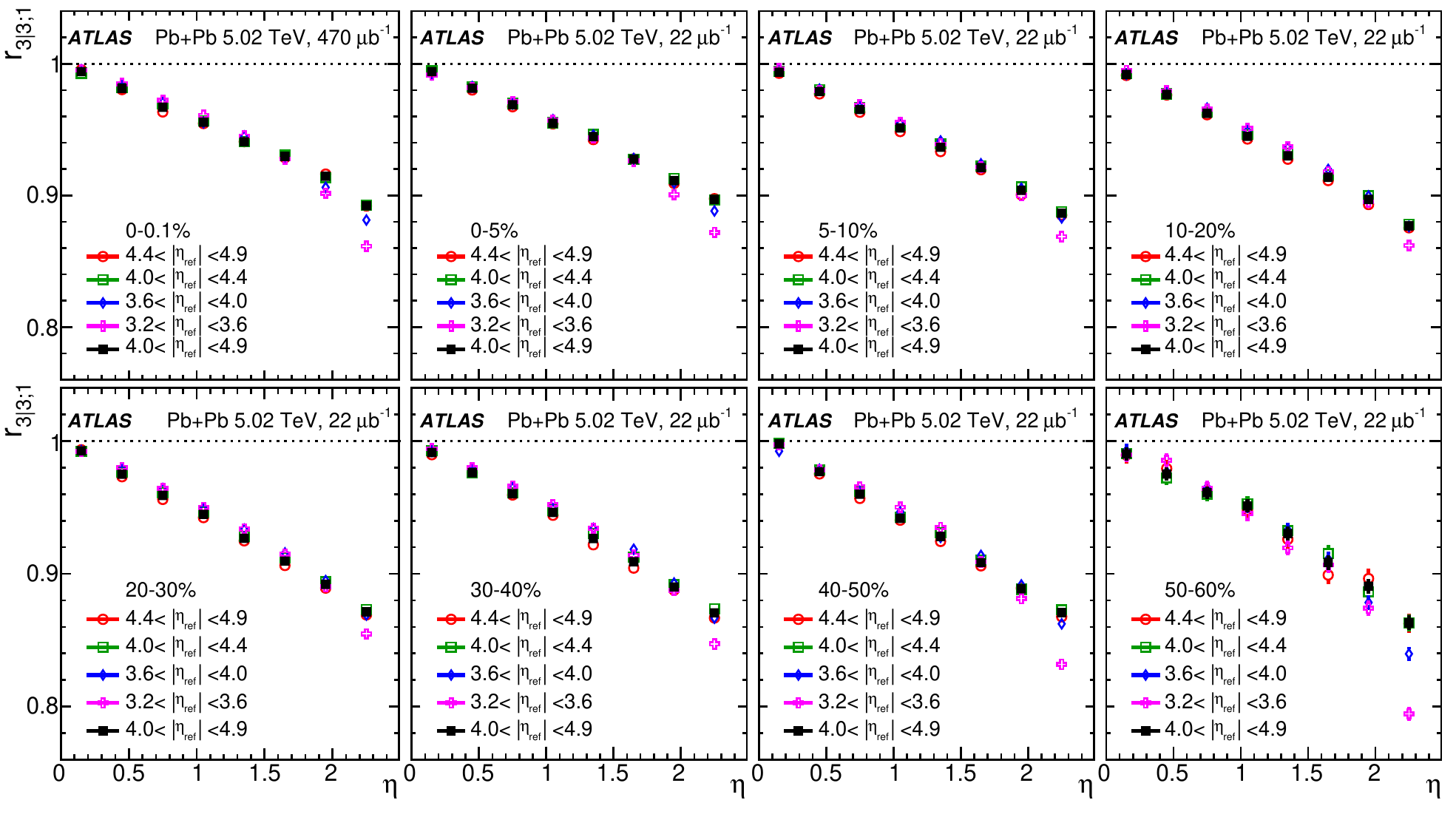}
\end{center}\vspace*{-0.3cm}
\caption{\label{fig:2}  The $r_{3|3;1}(\eta)$ measured for several $\etaref$ ranges. Each panel shows the results for one centrality range. The error bars are statistical only.}
\end{figure}

Figures~\ref{fig:1} and \ref{fig:2} show the sensitivity of $r_{2|2;1}$ and $r_{3|3;1}$, respectively, to the choice of the range of $\etaref$. A smaller $\etaref$ value implies a smaller pseudorapidity gap between $\eta$ and $\etaref$. The values of $r_{n|n;1}$ generally decrease with decreasing $\etaref$, possibly reflecting the contributions from the dijet correlations~\cite{Aad:2012bu}. However, such contributions should be reduced in the most central collisions due to large charged-particle multiplicity and jet-quenching~\cite{Gyulassy:2003mc} effects. Therefore, the decrease of $r_{n|n;1}$ in the most central collisions may also reflect the $\etaref$ dependence of $F_{n;1}^{\textrm r}$, as defined in Eq.~\eqref{eq:3}.  In this analysis, the reference flow vector is calculated from $4.0<\etaref<4.9$, which reduces the effect of dijets and provides good statistical precision. For this choice of $\etaref$ range, $r_{2|2;1}$ and $r_{3|3;1}$ show a linear decrease as a function of $\eta$ in most centrality intervals, indicating a significant breakdown of factorisation.  A similar comparison for $r_{4|4;1}$ can be found in the Appendix.


\begin{figure}[h!]
\begin{center}
\includegraphics[width=1\linewidth]{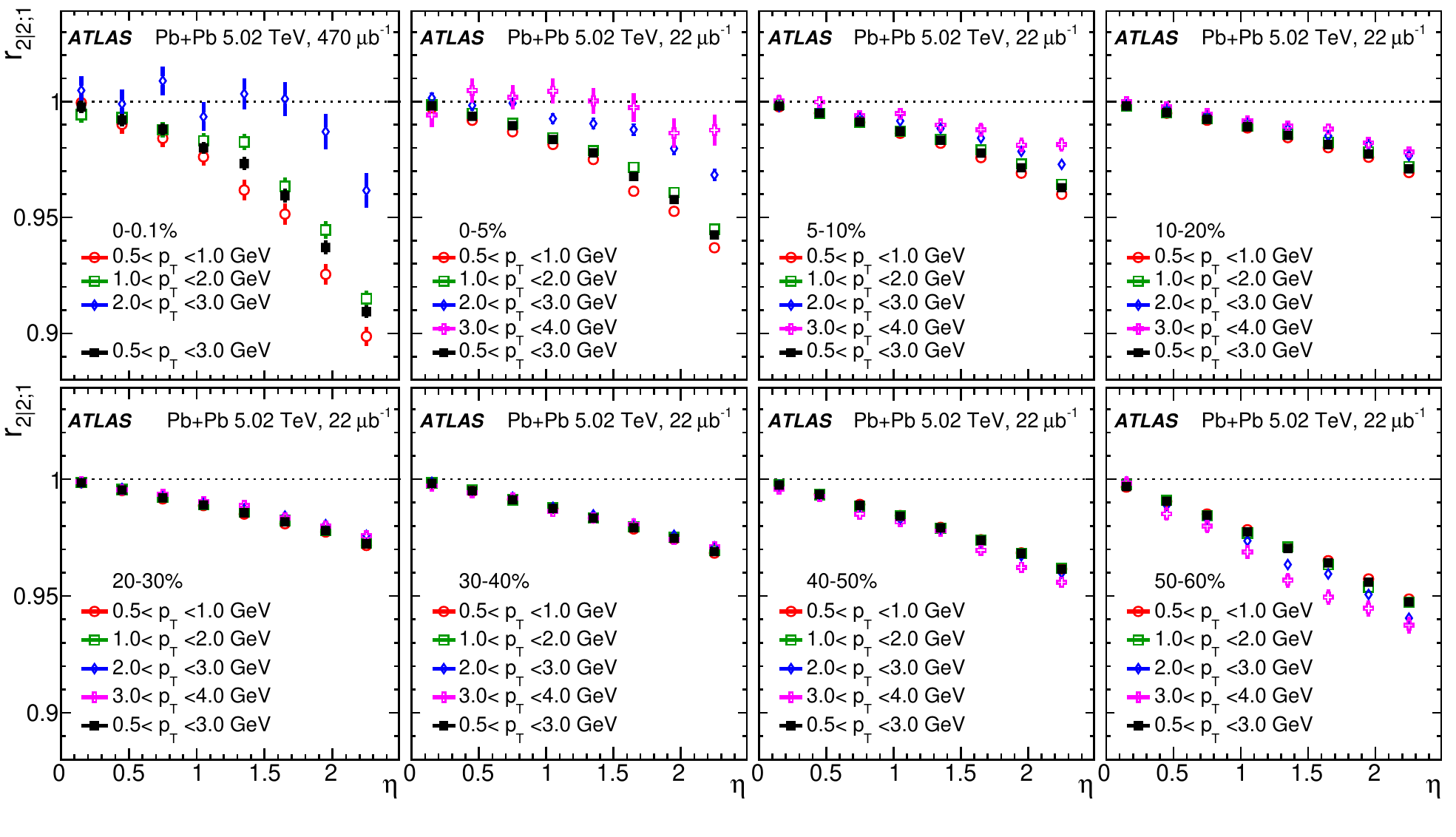}
\end{center}\vspace*{-0.3cm}
\caption{\label{fig:3} The $r_{2|2;1}(\eta)$ measured in several $\pT$ ranges. Each panel shows the results for one centrality range. The error bars are statistical only.}
\end{figure}

\begin{figure}[h!]
\begin{center}
\includegraphics[width=1\linewidth]{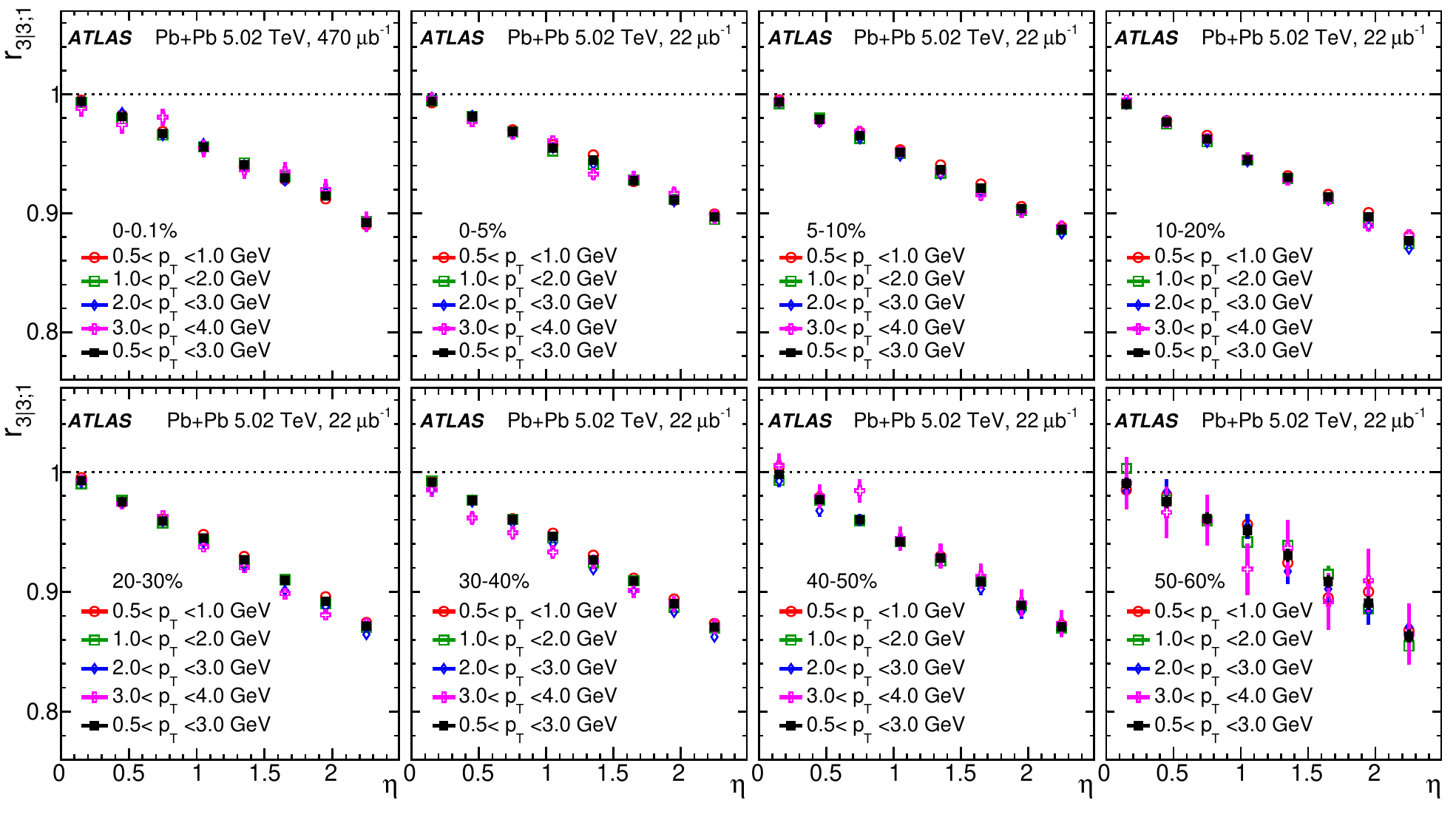}
\end{center}\vspace*{-0.3cm}
\caption{\label{fig:4} The $r_{3|3;1}(\eta)$ measured in several $\pT$ ranges. Each panel shows the results for one centrality range. The error bars are statistical only.}
\end{figure}

Figures~\ref{fig:3} and \ref{fig:4} show $r_{2|2;1}$ and $r_{3|3;1}$ calculated for several $\pT$ ranges of the charged particles in the ID. A similar comparison for $r_{4|4;1}$ can be found in the Appendix. If the longitudinal-flow asymmetry and twist reflect global properties of the event, the values of $r_{n|n;1}$ should not depend strongly on $\pT$. Indeed no dependence is observed, except for $r_{2|2;1}$ in the most central collisions and very peripheral collisions. The behaviour in central collisions may be related to the factorisation breaking of the $v_2$ as a function of $\pT$ and $\eta$~\cite{Aad:2012bu,Khachatryan:2015oea}. The behaviour in peripheral collisions is presumably due to increasing relative contributions from jets and dijets at higher $\pT$ and for peripheral collisions. Based on this, the measurements are performed using charged particles with $0.5<\pT<3$~GeV.
\section{Systematic uncertainties}
\label{sec:syscheck}
Since all observables are found to follow an approximately linear decrease with $\eta$, i.e.\ $D(\eta) \approx 1-c\eta$ for a given observable $D(\eta)$ where $c$ is a constant, the systematic uncertainty is presented as the relative uncertainty for $1-D(\eta)$ at $\eta=1.2$, the mid-point of the $\eta$ range. The systematic uncertainties in this analysis arise from event mixing, track selection, and reconstruction efficiency. Most of the systematic uncertainties enter the analysis through the particle weights in Eqs.~\eqref{eq:b1} and \eqref{eq:b2}. In general, the uncertainties for $r_{n|n;k}$ increase with $n$ and $k$, the uncertainties for $R_{n|n;2}$ increase with $n$, and all uncertainties are larger in the most central and more peripheral collisions. For $r_{2,3|2,3}$, $r_{2,2|4}$ and $r_{2,3|5}$, the uncertainties are significantly larger than for the other correlators. Each source is discussed separately below.

The effect of detector azimuthal non-uniformity is accounted for by the weight factor $d(\eta,\phi)$ in Eqs.~\eqref{eq:b1} and \eqref{eq:b2}. The effect of reweighting is studied by setting the weight to unity and repeating the analysis. The results are consistent with the default (weighted) results within statistical uncertainties, so no additional systematic uncertainty is included. Possible residual detector effects for each observable are further removed by subtracting those obtained from mixed events as described in Section~\ref{sec:corr}. Uncertainties due to the event-mixing procedure are estimated by varying the criteria for matching events in centrality and $z_{\mathrm{vtx}}$. The resulting uncertainty is in general found to be smaller than the statistical uncertainties. The event-mixing uncertainty for $r_{2|2;k}$ and $r_{3|3;k}$ is less than 1\% for $k=1$ and changes to about 0.4--8\% for $k=2$ and 0.6--10\% for $k=3$, while the uncertainty for $r_{4|4;1}$ and $r_{5|5;1}$ is in the range 1.5--3\% and 5--13\%, respectively. The uncertainty for $R_{n|n;2}$ is 1.5--6\% for $n=2$ and 3--14\% for $n=3$. The uncertainties for $r_{2,3|2,3}$, $r_{2,2|4}$ and $r_{2,2|5}$ are typically larger: 1--4\%, 1.5--16\% and 3--15\%.

The systematic uncertainty associated with the track quality selections is estimated by tightening or loosening the requirements on transverse impact parameter $|d_0|$ and longitudinal impact parameter $|z_0\sin\theta|$ used to select tracks. In each case, the tracking efficiency is re-evaluated and the analysis is repeated. The difference is observed to be larger in the most central collisions where the flow signal is smaller and the influence of falsely reconstructed tracks is higher. The difference is observed to be in the range 0.2--12\% for $r_{2|2;k}$ and $r_{3|3;k}$, 1.1--2\% for $r_{4|4;1}$,  3--6\% for $r_{5|5;1}$, 0.5--13\% for $R_{n|n;2}$, and 1--14\% for $r_{2,3|2,3}$, $r_{2,2|4}$ and $r_{2,2|5}$.

From previous measurements~\cite{Aad:2012bu,Chatrchyan:2013kba,Adam:2016ows}, the $v_n$ signal has been shown to have a strong dependence on $\pT$ but relatively weak dependence on $\eta$. Therefore, a $\pT$-dependent uncertainty in the track reconstruction efficiency $\epsilon(\eta,\pT)$ could affect the measured longitudinal flow correlation, through the particle weights. The uncertainty in the track reconstruction efficiency is due to differences in the detector conditions and known differences in the material between data and simulations. The uncertainty in the efficiency varies between 1\% and 4\%, depending on $\eta$ and $\pT$~\cite{Aad:2014vba}. The systematic uncertainty for each observable in Table~\ref{tab:obs} is evaluated by repeating the analysis with the tracking efficiency varied up and down by its corresponding uncertainty. For $r_{n|n;k}$ the uncertainties are in the range 0.1--2\%, depending on $n$ and $k$. For $R_{n|n;2}$ the uncertainties are in the range 0.1--1\%.  For $r_{2,3|2,3}$, $r_{2,2|4}$ and $r_{2,3|5}$, the uncertainties are in the range 0.1--2\%.

Due to the finite energy resolution and energy scale uncertainty of the FCal, the ${\bm q}_n(\etaref)$ calculated from the azimuthal distribution of the $\eT$ via Eqs.~\eqref{eq:1} and \eqref{eq:b2} differs from the true azimuthal distribution. However, since ${\bm q}_n(\etaref)$ appears in both the numerator and the denominator of the correlators studied in this paper, most of the effects associated with the FCal $\eT$ response are expected to cancel out. Two cross-checks are also performed to study the influence of the FCal response. In the first cross-check, only the FCal towers with $\eT$ above the $50^{\mathrm{th}}$ percentile are used to calculate the ${\bm q}_n(\etaref)$. The $|{\bm q}_n(\etaref)|$ value is different from the default analysis, but the values of the correlators are found to be consistent. In the second cross-check, HIJING events with imposed flow (see Section~\ref{sec:sel}) are used to study the FCal response. The ${\bm q}_n(\etaref)$ is calculated using both the generated $\eT$ and the reconstructed $\eT$, and the resulting correlators are compared with each other. The results are found to be consistent. Accordingly, no additional systematic uncertainty is added for the FCal response.

The systematic uncertainties from the different sources described above are added in quadrature to give the total systematic uncertainty for each observable. They are listed in Tables~\ref{tab:1}--\ref{tab:3}.
\begin{table}[!h]
\centering
\caption{\label{tab:1} Systematic uncertainties in percent for $1-r_{2|2;k}$ and $1-r_{3|3;k}$ at $\eta=1.2$ in selected centrality intervals.}
\small{\begin{tabular}{|c|ccc|ccc|ccc|}\hline 
                &\multicolumn{3}{|c|}{1$-r_{2|2;1}$}&\multicolumn{3}{c|}{1$-r_{2|2;2}$}&\multicolumn{3}{c|}{1$-r_{2|2;3}$}\\\hline
                & 0--5\%& 20--30\%&40--50\%& 0--5\%& 20--30\%&40--50\%& 0--5\%& 20--30\%&40--50\%\\\hline
Event mixing[\%]    & 0.8 & 0.2 & 0.3& 2.2 & 0.4 & 0.6& 6.0 & 0.6 & 2.1\\
Track selections[\%] &0.4 & 0.3 & 0.2& 1.5 & 0.4 & 0.9& 9.4 & 1.0 & 2.4\\
Reco. efficiency[\%]&0.3 & 0.1 & 0.1& 0.4 & 0.1 & 0.1&  0.9 & 0.1 & 0.1\\
Total[\%]           & 1.0 & 0.4 & 0.4& 2.7 & 0.6&1.1& 12 & 1.2 & 3.2\\\hline
                &\multicolumn{3}{|c|}{1$-r_{3|3;1}$}&\multicolumn{3}{c|}{1$-r_{3|3;2}$}&\multicolumn{3}{c|}{1$-r_{3|3;3}$}\\\hline
                & 0--5\%& 20--30\%&40--50\%& 0--5\%& 20--30\%&40--50\%& 0--5\%& 20--30\%&\\\hline
Event mixing[\%]    & 0.6 & 0.4 & 0.9 & 2.2 & 1.2 & 7.9 & 7.0 & 9.5 & \\
Track selections[\%] & 0.6 & 0.2 & 0.6 & 2.5 & 0.7 & 4.4 & 12  & 10  & \\
Reco. efficiency[\%]& 0.1 & 0.1 & 0.1 & 0.4 & 0.2 & 0.9 & 1.1 & 1.5 & \\
Total[\%]           & 0.9 & 0.5 & 1.1 & 3.4 & 1.5 & 9.1  & 14  & 14  & \\\hline
\end{tabular}}\normalsize
\end{table}

\begin{table}[!h]
\centering
\caption{\label{tab:2}Systematic uncertainties in percent for $1-R_{2|2;2}$,  $1-R_{3|3;2}$, $1-r_{4|4;1}$ and $1-r_{5|5;1}$ at $\eta=1.2$ in selected centrality intervals.}
\small{\begin{tabular}{|c|ccc|ccc|}\hline 
                &\multicolumn{3}{|c|}{1$-R_{2|2;2}$}&\multicolumn{3}{c|}{1$-R_{3|3;2}$}\\\hline
                &0--5\%& 20--30\%&40--50\%& 0--5\%& 20--30\%&40--50\%\\\hline
Event mixing    [\%]&6.1 & 1.5 & 1.5 & 4.6 & 2.9 & 14\\
Track selections [\%]&3.5 & 0.4 & 0.7 & 2.0 & 3.2 & 13\\
Reco. efficiency[\%]&0.2 & 0.1 & 0.1 & 0.1 & 0.2 & 0.5\\
Total           [\%]&7.1 & 1.6 & 1.7 & 5.1 & 4.4 & 20\\\hline
                &\multicolumn{3}{|c|}{1$-r_{4|4;1}$}&\multicolumn{3}{c|}{1$-r_{5|5;1}$}\\\hline
                &0--5\%& 20--30\%&40--50\%& 0--5\%& 20--30\%&40--50\%\\\hline
Event mixing    [\%]&1.8 & 1.5 & 2.7& 13 & 5.1 & 9.8\\
Track selections [\%]&1.5 & 1.1 & 2.0& 6.3 & 3.6 & 4.6\\
Reco. efficiency[\%]&0.3 & 0.3 & 0.6& 2.2 & 1.6 & 1.3\\
Total           [\%]&2.4 & 1.9 & 3.5& 15 & 6.5 & 11\\\hline
\end{tabular}}\normalsize
\end{table}
\begin{table}[!h]
\centering
\caption{\label{tab:3} Systematic uncertainties in percent for $1-r_{2,3|2,3}$, $1-r_{2,2|4}$ and $1-r_{2,3|5}$ at $\eta=1.2$ in selected centrality intervals.}
\small{\begin{tabular}{|c|ccc|ccc|ccc|}\hline 
                &\multicolumn{3}{|c|}{1$-r_{2,3|2,3}$}&\multicolumn{3}{c|}{1$-r_{2,2|4}$}&\multicolumn{3}{c|}{1$-r_{2,3|5}$}\\\hline
                &0--5\% & 20--30\% &40--50\% & 0--5\% & 20--30\% &40--50\%& 0--5\% & 20--30\% &40--50\%\\\hline
Event mixing    [\%]& 4.1 & 1.7 & 3.2   & 16  & 1.5 & 2.4   & 15  & 3.4 & 7.8\\
Track selections [\%]& 1.4 & 0.5 & 2.0   & 12  & 1.6 & 1.5   & 14  & 2.0 & 7.4\\
Reco. efficiency[\%]& 0.1 & 0.0 & 0.1   & 1.6 & 0.1 & 0.1   & 1.2 & 0.1 & 0.5\\
Total           [\%]& 4.4 & 1.8 & 3.8   & 21  & 2.2 & 2.9   & 21  & 4.0 & 11\\\hline
\end{tabular}}\normalsize
\end{table}

\section{Results}
\label{sec:res}
The presentation of the results is structured as follows. Section~\ref{sec:res1} presents the results for $r_{n|n;1}$ and $R_{n|n;2}$ and the comparison between the two collision energies. Section~\ref{sec:res2} shows the results for $r_{n|n;k}$ for $k>1$. The scaling relation from Eq.~\eqref{eq:5} is tested and the contributions from $v_n$ FB asymmetry and event-plane twist are estimated. Results for the mixed-harmonic correlators, Eqs.~\eqref{eq:8}--\eqref{eq:10}, are presented in Section~\ref{sec:res3} and checked for compatibility with the hydrodynamical picture. The measurements are performed using charged particles with $0.5<\pT<3$ GeV, and the reference flow vector is calculated with $4.0<|\etaref|<4.9$. Most results are shown for the $\sqrtsnn=5.02$~TeV Pb+Pb dataset, which has better statistical precision. The results for the $\sqrtsnn=2.76$~TeV Pb+Pb dataset are shown only for $r_{n|n;1}$ and $R_{n|n;2}$. 

\subsection{$r_{n|n;1}$ and $R_{n|n;2}$ at two collision energies}
\label{sec:res1}
\begin{figure}[h!]
\begin{center}
\includegraphics[width=0.8\linewidth]{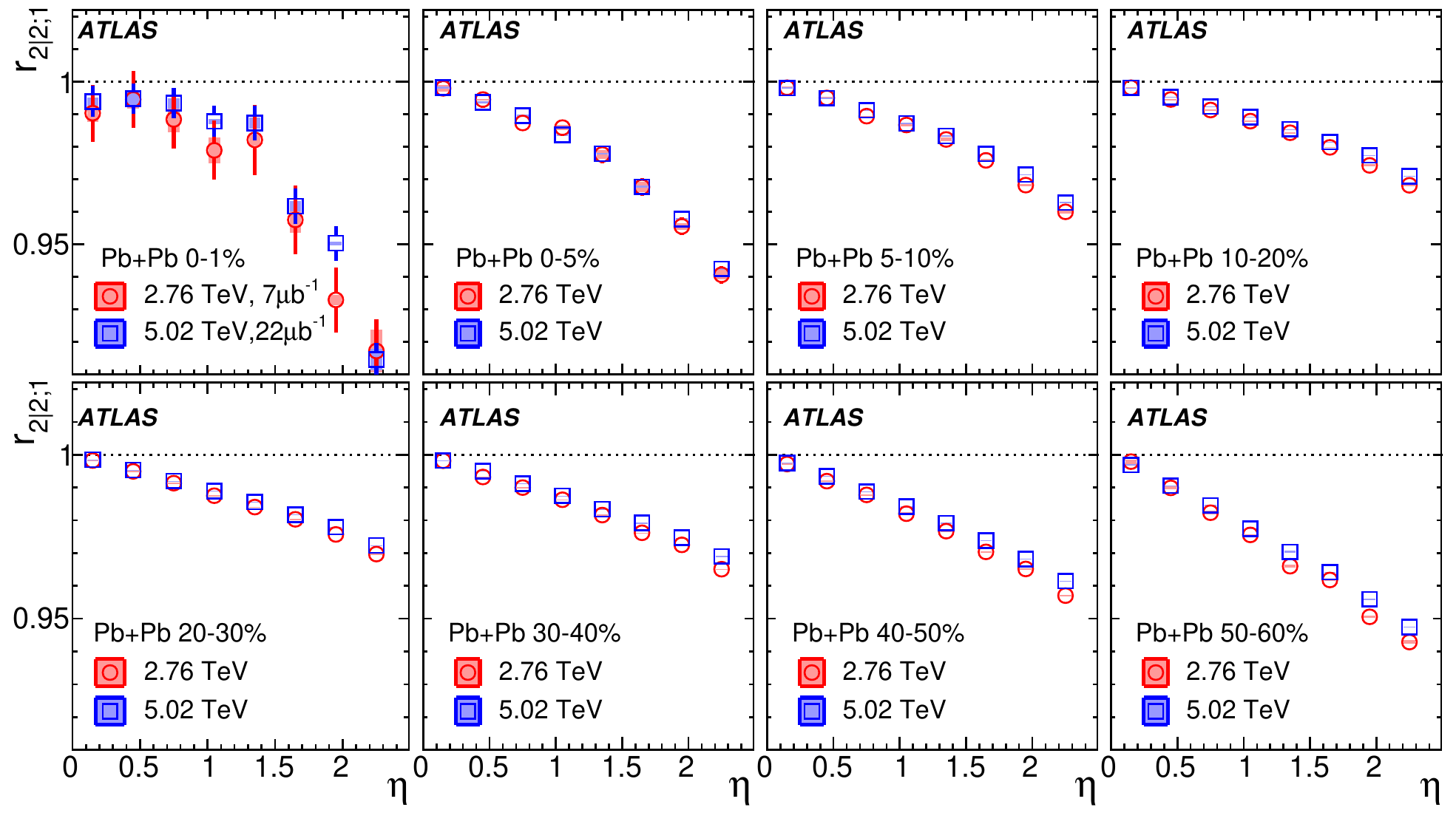}
\end{center}
\caption{\label{fig:5} The $r_{2|2;1}(\eta)$ compared between the two collision energies. Each panel shows results from one centrality interval. The error bars and shaded boxes are statistical and systematic uncertainties, respectively.}
\end{figure}
Figure~\ref{fig:5} shows $r_{2|2;1}$ in various centrality intervals at the two collision energies. The correlator shows a linear decrease with $\eta$, except in the most central collisions. The decreasing trend is weakest around the 20--30\% centrality range, and is more pronounced in both more central and more peripheral collisions. This centrality dependence is the result of a strong centrality dependence of the $v_2$ associated with the average elliptic geometry~\cite{Huo:2013qma}. The decreasing trend at $\sqrtsnn=2.76$~TeV is slightly stronger than that at \mbox{$\sqrtsnn=5.02$~TeV}, which is expected as the collision system becomes less boost-invariant at lower collision energy~\cite{Pang:2015zrq}.

\begin{figure}[h!]
\begin{center}
\includegraphics[width=0.8\linewidth]{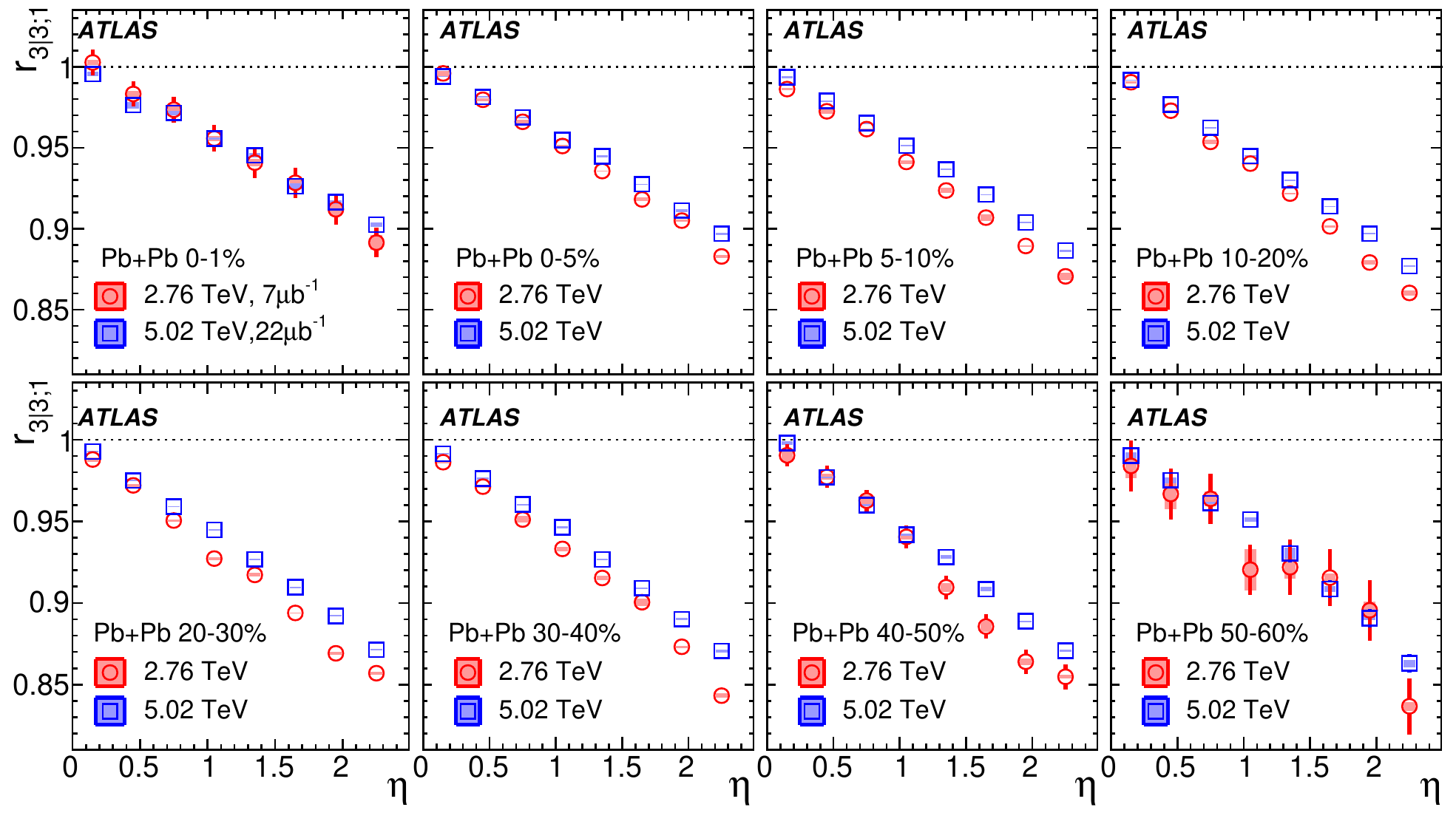}
\end{center}\vspace*{-0.7cm}
\caption{\label{fig:6}  The $r_{3|3;1}(\eta)$ compared between the two collision energies. Each panel shows results from one centrality interval. The error bars and shaded boxes are statistical and systematic uncertainties, respectively.}
\end{figure}

\begin{figure}[h!]
\begin{center}
\includegraphics[width=0.8\linewidth]{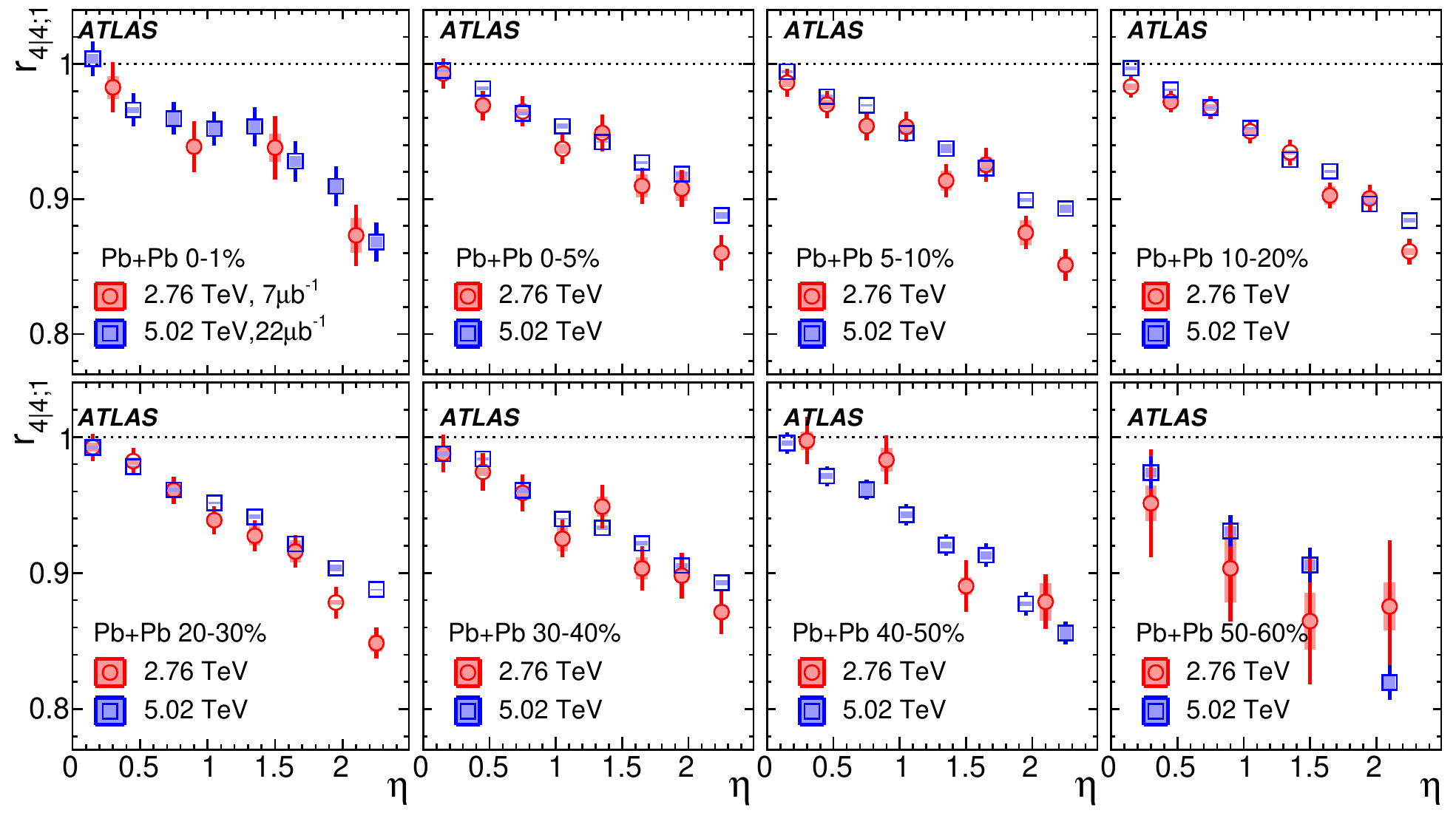}
\end{center}\vspace*{-0.7cm}
\caption{\label{fig:7}  The $r_{4|4;1}(\eta)$ compared between the two collision energies. Each panel shows results from one centrality interval. The error bars and shaded boxes are statistical and systematic uncertainties, respectively.}
\end{figure}

Figures~\ref{fig:6} and \ref{fig:7} show the results for $r_{3|3;1}$ and $r_{4|4;1}$, respectively, at the two collision energies.
A linear decrease as a function of $\eta$ is observed for both correlators, and the rate of the decrease is approximately independent of centrality. This centrality independence could be due to the fact that $v_3$ and $v_4$ are driven mainly by fluctuations in the initial state. The rate of the decrease is also observed to be slightly stronger at lower collision energy.

The decreasing trend of $r_{n|n;1}$ for $n=2$--4 in Figures~\ref{fig:5}--\ref{fig:7} indicates significant breakdown of the factorisation of two-particle flow harmonics into those between different $\eta$ ranges. However, the size of the factorisation breakdown depends on the harmonic order $n$, collision centrality, and collision energy. The results have also been compared with those from the CMS Collaboration~\cite{Khachatryan:2015oea}, with the $\etaref$ chosen to be $4.4<|\etaref|<4.9$ to match the CMS choice of $\etaref$. The two results agree very well with each other, and details are shown in the Appendix.

Figures~\ref{fig:8} and \ref{fig:9} show $R_{2|2;2}$ and $R_{3|3;2}$ in several centrality intervals. Both observables follow a linear decrease with $\eta$ and the decreasing trends are stronger at lower collision energy.
\begin{figure}[h!]
\begin{center}
\includegraphics[width=0.71\linewidth]{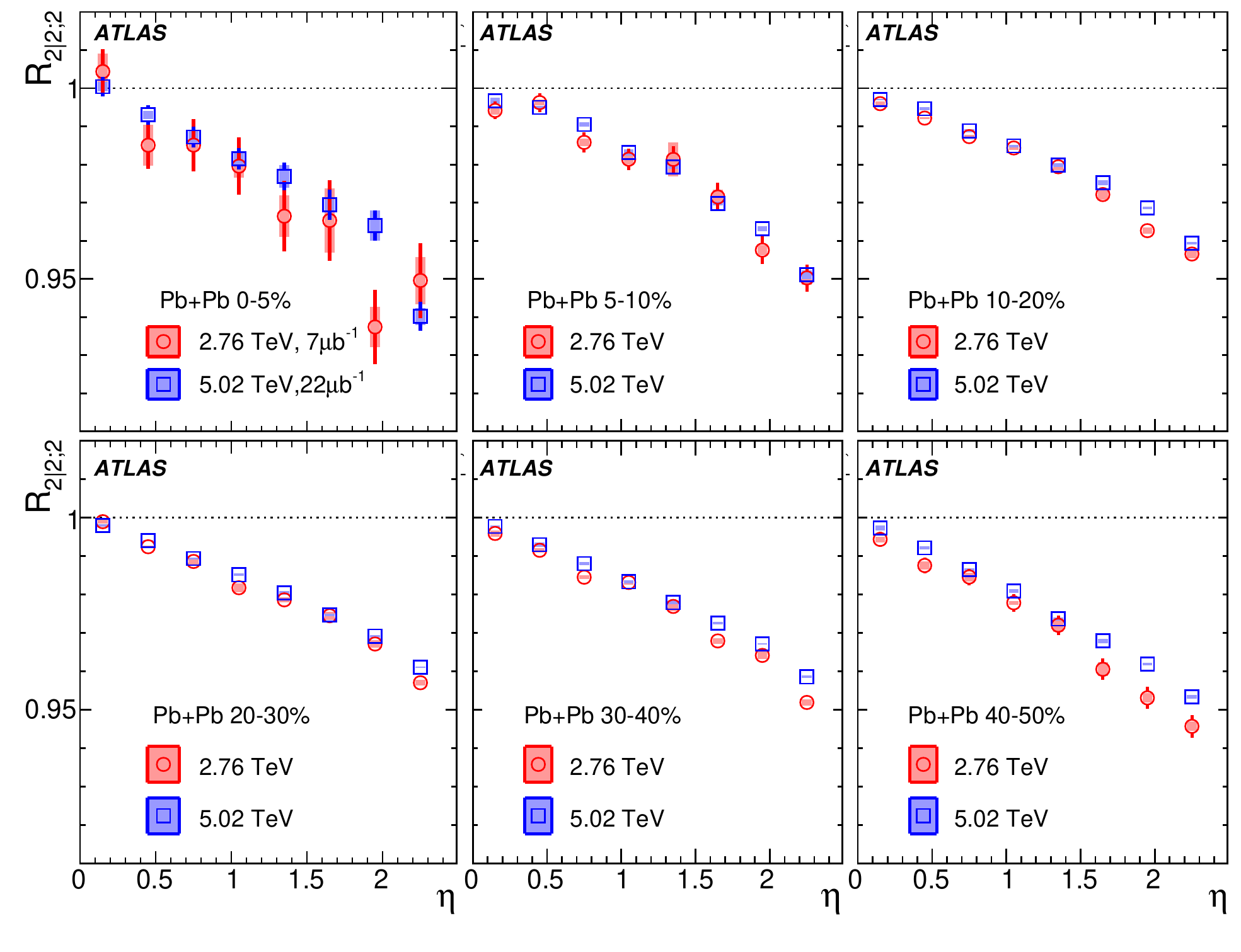}
\end{center}\vspace*{-0.3cm}
\caption{\label{fig:8}  The $R_{2|2;2}(\eta)$ compared between the two collision energies. Each panel shows results from one centrality interval. The error bars and shaded boxes are statistical and systematic uncertainties, respectively.}
\end{figure}
\begin{figure}[h!]
\begin{center}
\includegraphics[width=0.73\linewidth]{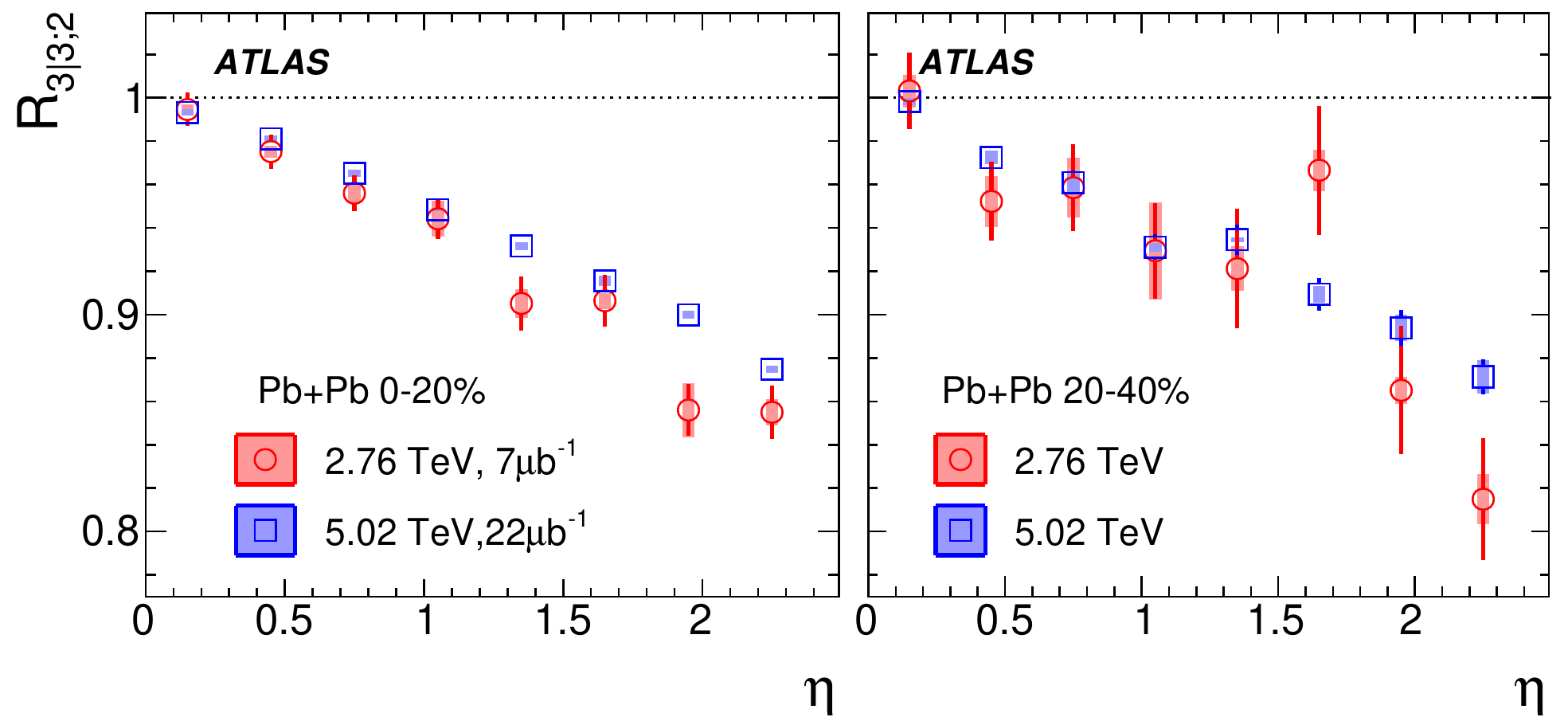}
\end{center}\vspace*{-0.3cm}
\caption{\label{fig:9}  The $R_{3|3;2}(\eta)$ compared between the two collision energies. Each panel shows results from one centrality interval. The error bars and shaded boxes are statistical and systematic uncertainties, respectively.}
\end{figure}

The measured $r_{n|n;k}$ and $R_{n|n;2}$ are parameterised with linear functions,
\begin{eqnarray}
\label{eq:20}
r_{n|n;k} = 1-2F_{n;k}^{\textrm r}\;\eta\;,\;\;R_{n|n;2} = 1-2F_{n;2}^{\textrm R}\;\eta\;,
\end{eqnarray}
where the slope parameters are calculated as linear-regression coefficients,
\begin{eqnarray}
\label{eq:20b}
 F_{n;k}^{\textrm r}=\frac{\sum_i(1-r_{n|n;k}(\eta_i))\eta_i}{2\sum_i\eta_i^2}\;,\;\; F_{n;2}^{\textrm R}=\frac{\sum_i(1-R_{n|n;2}(\eta_i))\eta_i}{2\sum_i\eta_i^2}\;,
\end{eqnarray}
which characterise the average $\eta$-weighted deviation of $r_{n|n;1}(\eta)$ and $R_{n|n;2}(\eta)$ from unity. The sum runs over all data points. If $r_{n|n;k}$ and $R_{n|n;2}$ are a linear function in $\eta$, the linear-regression coefficients are equivalent to a fit to Eq.~\eqref{eq:20}. However, these coefficients are well defined even if the observables have significant nonlinear behaviour, which is the case for $r_{2|2;k}$ and $R_{2|2;2}$ in the 0--20\% centrality range. 

\begin{figure}[h!]
\begin{center}
\includegraphics[width=0.88\linewidth]{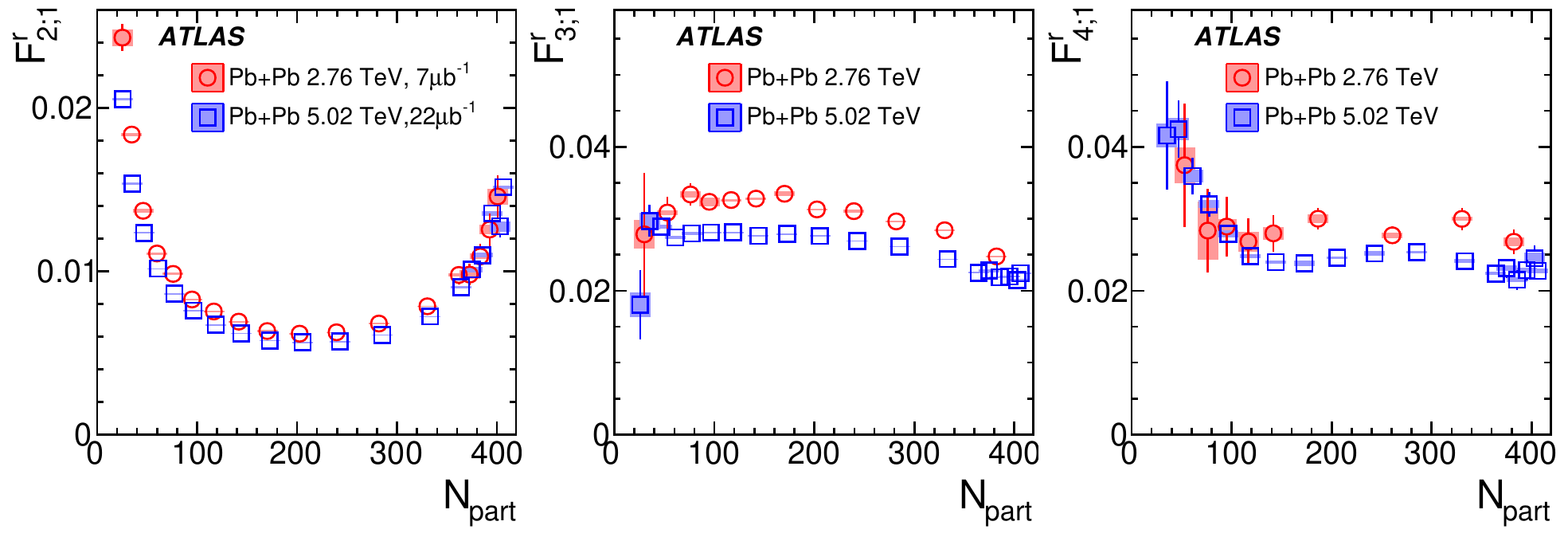}
\end{center}\vspace*{-0cm}
\caption{\label{fig:10} Centrality dependence of $F_{2;1}^{\textrm r}$ (left panel), $F_{3;1}^{\textrm r}$ (middle panel) and $F_{4;1}^{\textrm r}$ (right panel) for Pb+Pb at 2.76~TeV (circles) and 5.02~TeV (squares). The error bars and shaded boxes are statistical and systematic uncertainties, respectively. The widths of the centrality intervals are not fixed but are optimised to reduce the uncertainty.}
\end{figure}

\begin{figure}[h!]
\begin{center}
\includegraphics[width=0.88\linewidth]{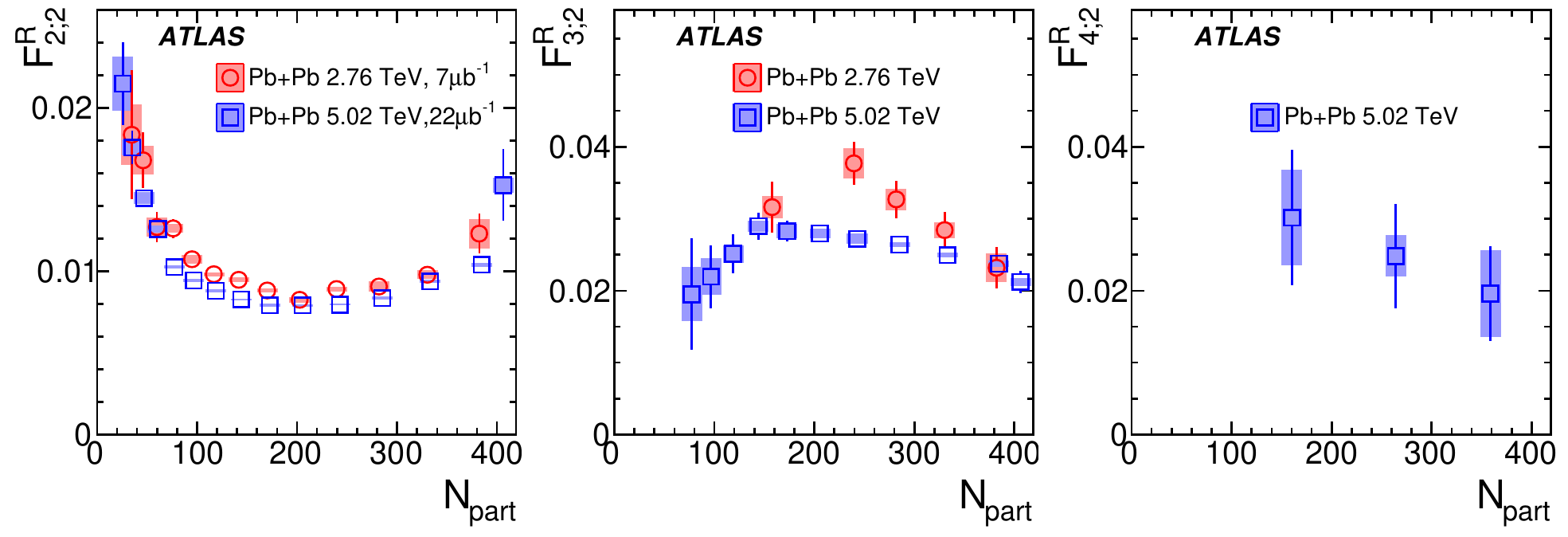}
\end{center}\vspace*{-0.2cm}
\caption{\label{fig:11} Centrality dependence of $F_{2;2}^{\textrm R}$ (left panel), $F_{3;2}^{\textrm R}$ (middle panel) and $F_{4;2}^{\textrm R}$ (right panel) for Pb+Pb at 2.76~TeV (circles) and 5.02~TeV (squares). The error bars and shaded boxes are statistical and systematic uncertainties, respectively. The widths of the centrality intervals are not fixed but are optimised to reduce the uncertainty.}
\end{figure}

\begin{figure}[h!]
\begin{center}
\includegraphics[width=0.9\linewidth]{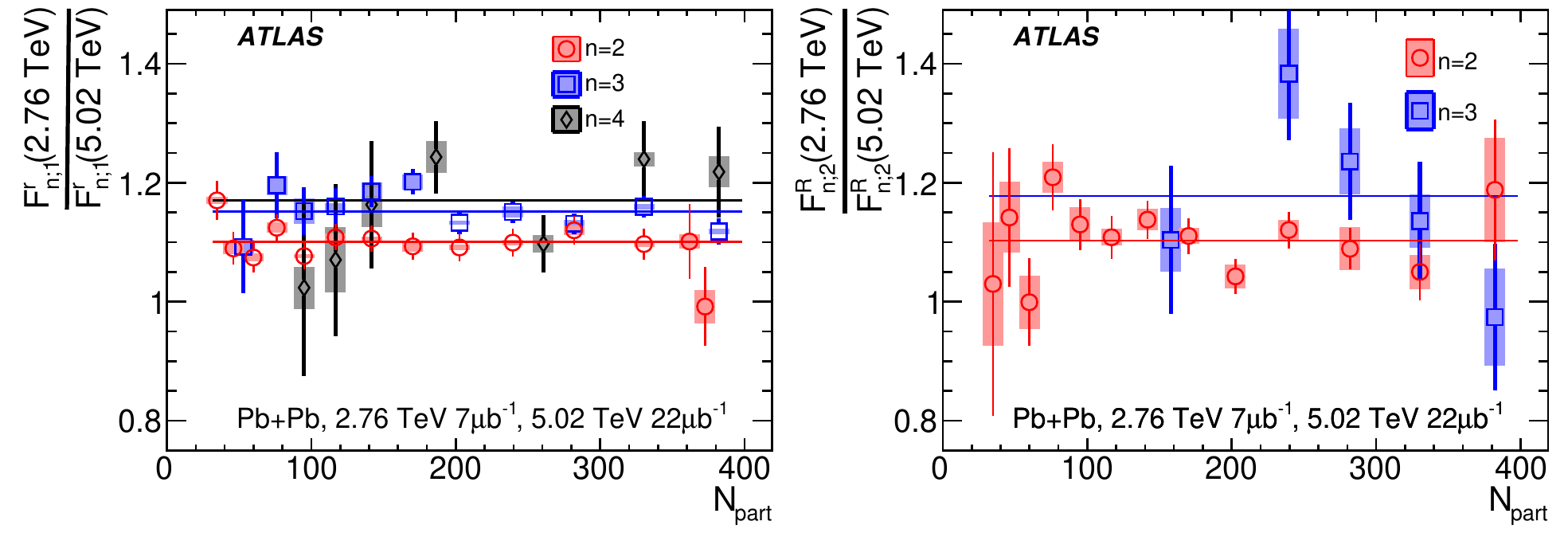}
\end{center}\vspace*{-0.4cm}
\caption{\label{fig:11b} Centrality dependence of ratio of $F_{n;1}^{\textrm r}$ values (left panel) and $F_{n;2}^{\textrm R}$ values (right panel) at 2.76~TeV and 5.02~TeV. The lines indicate the average values in the range $30<\npart<400$, with the results and fit uncertainties given by Table~\ref{tab:fit}. The error bars and shaded boxes are statistical and systematic uncertainties, respectively. }
\end{figure}

The extracted slope parameters $F_{n;1}^{\textrm r}$ and $F_{n;2}^{\textrm R}$ are plotted as a function of centrality in terms of $\npart$, in Figures~\ref{fig:10} and \ref{fig:11}, respectively. The values of $F_{2;1}^{\textrm r}$ and $F_{2;2}^{\textrm R}$ first decrease and then increase as a function of increasing $\npart$. The larger values in central and peripheral collisions are related to the fact that $v_2$ is more dominated by the initial geometry fluctuations. The slopes for higher-order harmonics are significantly larger. As a function of $\npart$, a slight decrease in $F_{3;1}^{\textrm r}$ and $F_{3;2}^{\textrm R}$ is observed for $\npart>200$, as well as an increase in $F_{4;1}^{\textrm r}$ for $\npart<100$. The values of $F_{n;1}^{\textrm r}$ and $F_{n;2}^{\textrm R}$ are larger with decreasing $\sqrtsnn$, as the rapidity profile of the initial state is more compressed due to smaller beam rapidity $y_{\mathrm {beam}}$ at lower $\sqrtsnn$. This energy dependence has been predicted for $F_{n;1}^{\textrm r}$ in hydrodynamic model calculations~\cite{Pang:2015zrq}, and it is quantified in Figure~\ref{fig:11b} via the ratio of $F_{2;1}^{\textrm r}$ values and of $F_{2;2}^{\textrm R}$ values at the two energies. The weighted averages of the ratios calculated in the range $30<\npart<400$ are given in Table~\ref{tab:fit}. Compared to $\sqrtsnn=5.02$~TeV, the values of $F_{2;1}^{\textrm r}$ and $F_{2;2}^{\textrm R}$ at $\sqrtsnn=2.76$~TeV are about 10\% higher, and the values of $F_{3;1}^{\textrm r}$ and $F_{4;1}^{\textrm r}$ are about 16\% higher. 

\begin{figure}[h!]
\begin{center}
\includegraphics[width=0.9\linewidth]{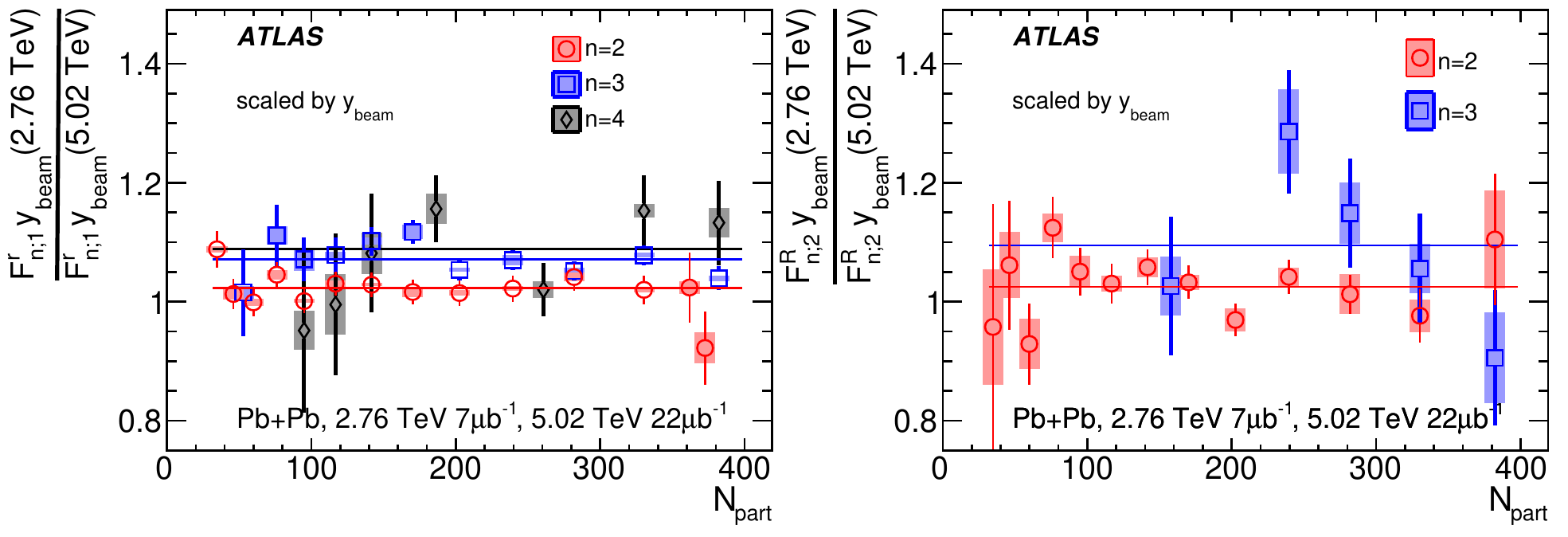}
\end{center}\vspace*{-0.4cm}
\caption{\label{fig:11c} Centrality dependence of ratio of $\hat{F}_{n;1}^{\textrm r} \equiv F_{n;1}^{\textrm r}y_{\mathrm {beam}}$ values (left panel) and $\hat{F}_{n;2}^{\textrm R}\equiv F_{n;2}^{\textrm R}y_{\mathrm {beam}}$ values (right panel) at 2.76~TeV and 5.02~TeV. The lines indicate the average values in the range $30<\npart<400$, with the results and fit uncertainties given by Table~\ref{tab:fit}. The error bars and shaded boxes are statistical and systematic uncertainties, respectively. }
\end{figure}

If the change of correlators with $\sqrtsnn$ were entirely due to the change of $y_{\mathrm {beam}}$, then the correlators would be expected to follow a universal curve when they are rescaled by $y_{\mathrm {beam}}$, i. e. $r_{n|n;k}(\eta/y_{\mathrm {beam}})$ and $R_{n|n;2}(\eta/y_{\mathrm {beam}})$ should not depend on $\sqrtsnn$. In this case, the slopes parameters multiplified by the beam rapidity, $\hat{F}_{n;1}^{\textrm r}\equiv F_{n;1}^{\textrm r}y_{\mathrm {beam}}$ and $\hat{F}_{n;2}^{\textrm R}\equiv F_{n;2}^{\textrm R}y_{\mathrm {beam}}$, should not depend on $\sqrtsnn$. The beam rapidity is $y_{\mathrm {beam}}=7.92$ and 8.52 for $\sqrtsnn=2.76$ and 5.02 TeV, respectively, which leads to a 7.5\% reduction in the ratio. Figure~\ref{fig:11c} shows the ratio of $\hat{F}_{2;1}^{\textrm r}$ values and of $\hat{F}_{2;2}^{\textrm R}$ values at the two energies, and the weighted averages of the ratios calculated in the range $30<\npart<400$ are given in Table~\ref{tab:fit}. The $y_{\mathrm {beam}}$-scaling accounts for a large part of the $\sqrtsnn$ dependence.  Compared to $\sqrtsnn=5.02$~TeV, the values of $\hat{F}_{2;1}^{\textrm r}$ and $\hat{F}_{2;2}^{\textrm R}$ at $\sqrtsnn=2.76$~TeV are about 3\% higher, and the values of $\hat{F}_{3;1}^{\textrm r}$ and $\hat{F}_{4;1}^{\textrm r}$ are about 8\% higher, so this level of difference remains after accounting for the change in the beam rapidity. 


\begin{table}[!h]
\centering
\caption{\label{tab:fit} Results of the fits to the ratio of $F_{n;1}^{\textrm r}$,  $F_{n;2}^{\textrm R}$, $\hat{F}_{n;1}^{\textrm r}\equiv F_{n;1}^{\textrm r}y_{\mathrm {beam}}$ and  $\hat{F}_{n;2}^{\textrm R}\equiv F_{n;2}^{\textrm R}y_{\mathrm {beam}}$ values at the two energies in the range $30<\npart<400$ shown in Figures~\ref{fig:11b} and \ref{fig:11c}. The uncertainties include both statistical and systematic uncertainties.}
\begin{tabular}{|c|ccc|}\hline 
                                                               &$n=2$          &$n=3$          & $n=4$\\\hline
\multirow{2}{*}{$F_{n;1}^{\textrm r}(2.76~{\textrm{TeV}})\vphantom{\int}{/}F_{n;1}^{\textrm r}(5.02~{\textrm{TeV}})$} & \multirow{2}{*}{$1.100\pm 0.010$}  &\multirow{2}{*}{$1.152\pm 0.011$} &\multirow{2}{*}{ $1.17\pm 0.036$}\\
&&&\\\hline
\multirow{2}{*}{$F_{n;2}^{\textrm R}(2.76~{\textrm{TeV}})\vphantom{\int}{/}F_{n;2}^{\textrm R}(5.02~{\textrm{TeV}})$} & \multirow{2}{*}{$1.103\pm 0.026$}  &\multirow{2}{*}{$1.18\pm 0.08$} &\multirow{2}{*}{--}\\
&&&\\\hline\hline
\multirow{2}{*}{$\hat{F}_{n;1}^{\textrm r}(2.76~{\textrm{TeV}})\vphantom{\int}{/}\hat{F}_{n;1}^{\textrm r}(5.02~{\textrm{TeV}})$} & \multirow{2}{*}{$1.023\pm 0.009$}  &\multirow{2}{*}{$1.071\pm 0.010$} &\multirow{2}{*}{ $1.088\pm 0.033$}\\
&&&\\\hline
\multirow{2}{*}{$\hat{F}_{n;2}^{\textrm R}(2.76~{\textrm{TeV}})\vphantom{\int}{/}\hat{F}_{n;2}^{\textrm R}(5.02~{\textrm{TeV}})$} & \multirow{2}{*}{$1.025\pm 0.024$}  &\multirow{2}{*}{$1.10\pm 0.07$} &\multirow{2}{*}{--}\\
&&&\\\hline
\end{tabular}
\end{table}

\vspace*{-0.5cm}\subsection{Higher-order moments}
\label{sec:res2}
The longitudinal correlations of higher-order moments of harmonic flow carry information about the EbyE flow fluctuations in pseudorapidity. In the simple model described in Ref.~\cite{Jia:2017kdq}, the decrease in $r_{n|n;k}$ is expected to scale with $k$ as given by Eq.~\eqref{eq:5}.

\begin{figure}[h!]
\begin{center}
\includegraphics[width=0.9\linewidth]{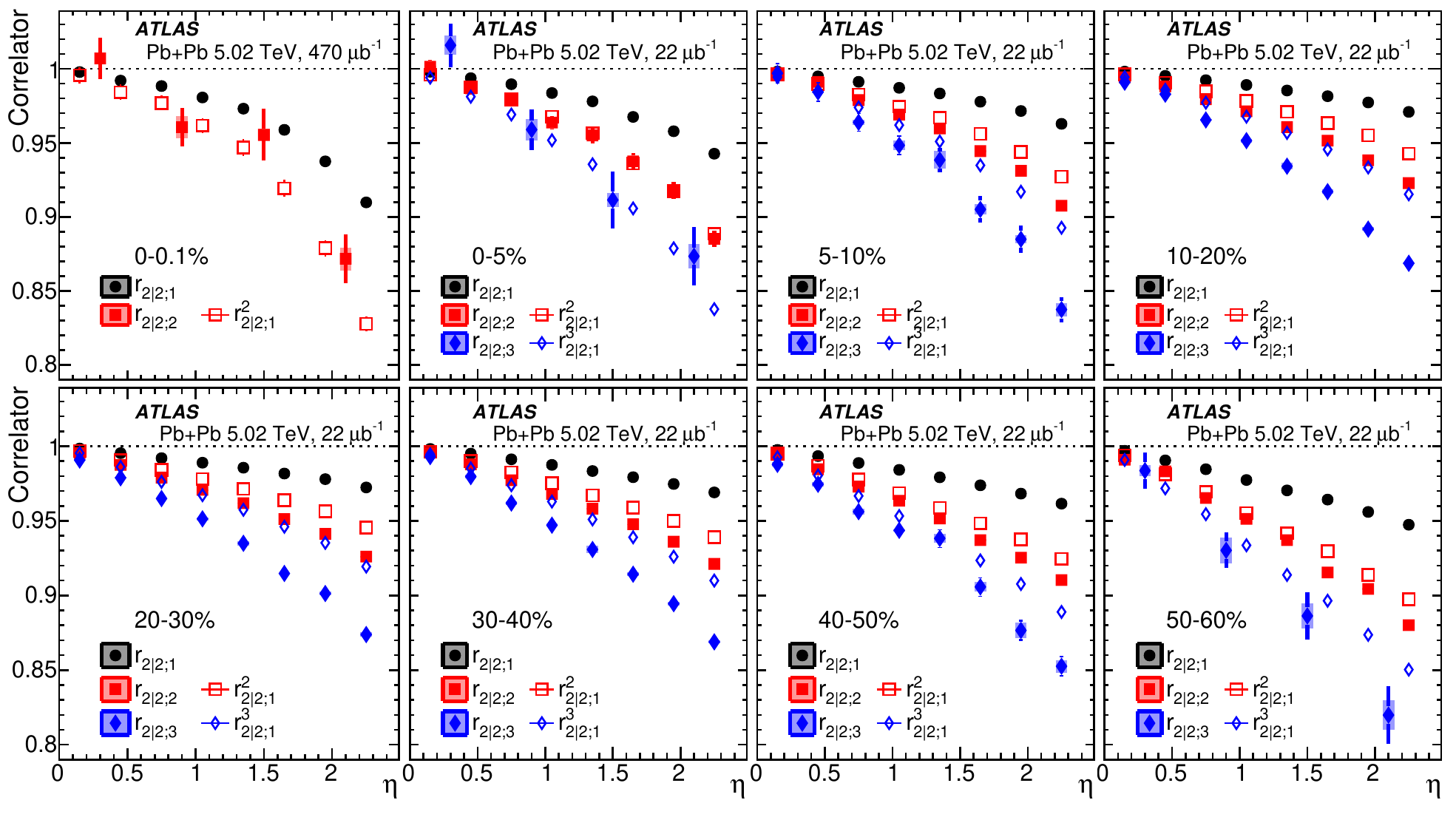}
\end{center}
\caption{\label{fig:12} The $r_{2|2;k}$ for $k=1$--3 compared with $r_{2|2;1}^k$ for $k=2$--3 in various centrality intervals for Pb+Pb collisions at 5.02~TeV. The error bars and shaded boxes are statistical and systematic uncertainties, respectively. The data points for $k=2$ or 3 in some centrality intervals are rebinned to reduce the uncertainty.}
\end{figure}

Figure~\ref{fig:12} compares the results for $r_{2|2;k}$ for $k=1$--3 (solid symbols)  with $r_{2|2;1}^k$ for $k=2$--3 (open symbols). The data follow the scaling relation from Eq.~\eqref{eq:5} in the most central collisions (0--5\% centrality) where $v_2$ is driven by the initial-state fluctuations. In other centrality intervals, where the average geometry is more important for $v_2$, the $r_{2|2;k}$ ($k=2$ and 3) data show stronger decreases with $\eta$ than $r_{2|2;1}^k$.

\begin{figure}[h!]
\begin{center}
\includegraphics[width=0.9\linewidth]{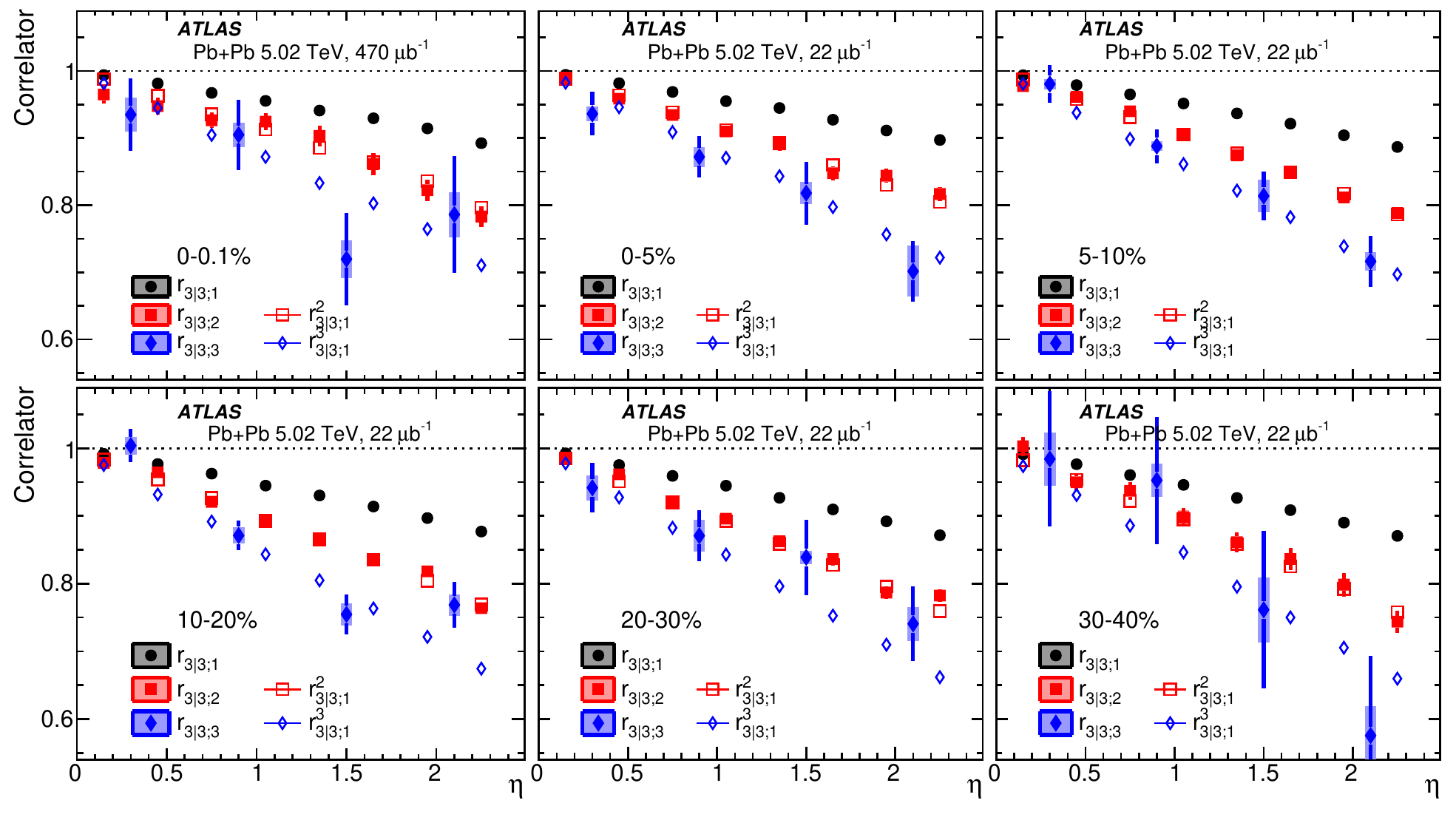}
\end{center}
\caption{\label{fig:13} The $r_{3|3;k}$ for $k=1$--3 compared with $r_{3|3;1}^k$ for $k=2$--3 in various centrality intervals for Pb+Pb collisions at 5.02~TeV. The error bars and shaded boxes are statistical and systematic uncertainties, respectively. The data points for $k=2$ or 3 in some centrality intervals are rebinned to reduce the uncertainty.}
\end{figure}

A similar study is performed for third-order harmonics, and the results are shown in Figure~\ref{fig:13}. The data follow approximately the scaling relation Eq.~\eqref{eq:5} in all centrality intervals. 

\begin{figure}[h!]
\begin{center}
\includegraphics[width=0.9\linewidth]{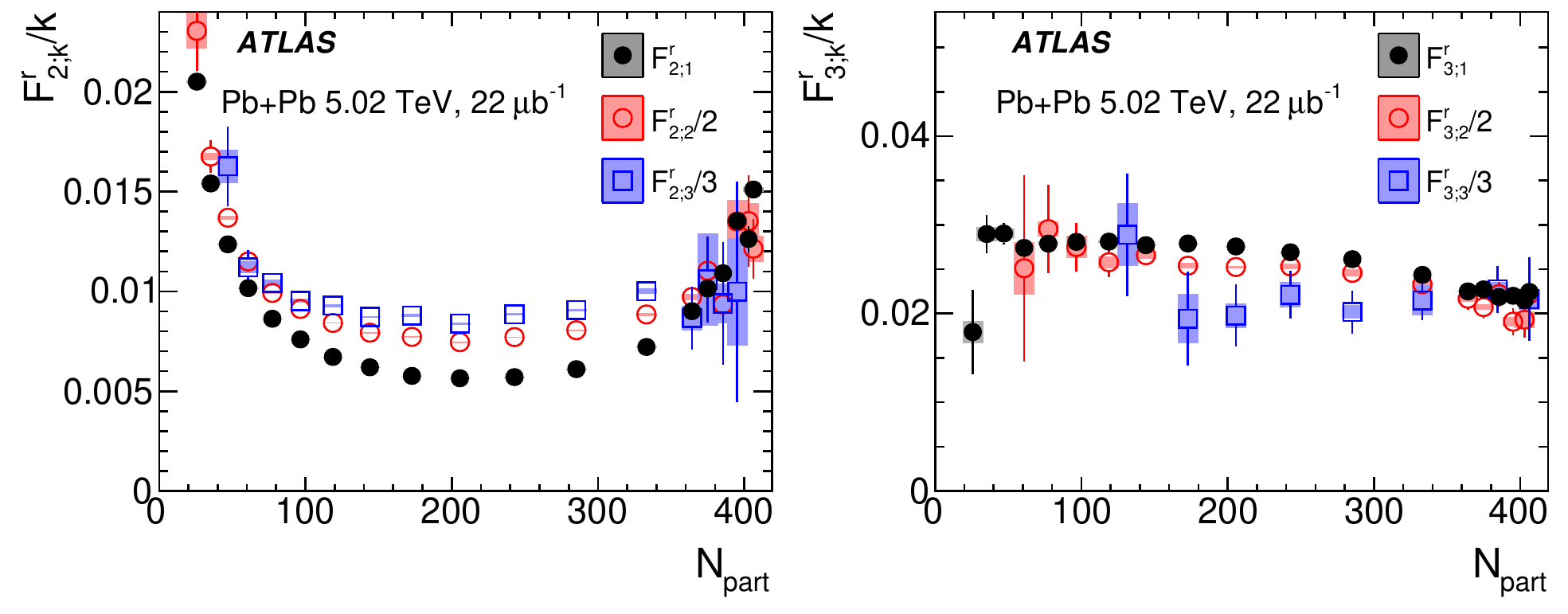}
\end{center}
\caption{\label{fig:14} The values of $F_{n;k}^{\textrm r}/k$ for $k=1$,2 and 3 for $n=2$ (left panel) and $n=3$ (right panel), respectively. The error bars and shaded boxes are statistical and systematic uncertainties, respectively. The widths of the centrality intervals are not fixed but are optimised to reduce the uncertainty.}
\end{figure}

To quantify the difference between $r_{n|n;k}$ and $r_{n|n;1}^k$, the slopes ($F_{n;k}^{\textrm r}$) of $r_{n|n;k}$ are calculated via Eqs.~\eqref{eq:20} and \eqref{eq:20b}. The scaled quantities, $F_{n;k}^{\textrm r}/k$, are then compared with each other as a function of centrality in Figure~\ref{fig:14}. For second-order harmonics, the data show clearly that over most of the centrality range $F_{2;3}^{\textrm r}/3>F_{2;2}^{\textrm r}/2>F_{2;1}^{\textrm r}$, implying  $F_{2;k}^{\textrm r}>kF_{2;1}^{\textrm r}$. However, for the most central and most peripheral collisions the quantities approach each other. On the other hand, a slightly opposite trend for the third-order harmonics, $F_{3;3}^{\textrm r}/3\lesssim F_{3;2}^{\textrm r}/2\lesssim F_{3;1}^{\textrm r}$, i.e.\ $F_{3;k}^{\textrm r}\lesssim kF_{3;1}^{\textrm r}$, is observed in mid-central collisions ($150<\npart<350$).

\begin{figure}[h!]
\begin{center}
\includegraphics[width=0.8\linewidth]{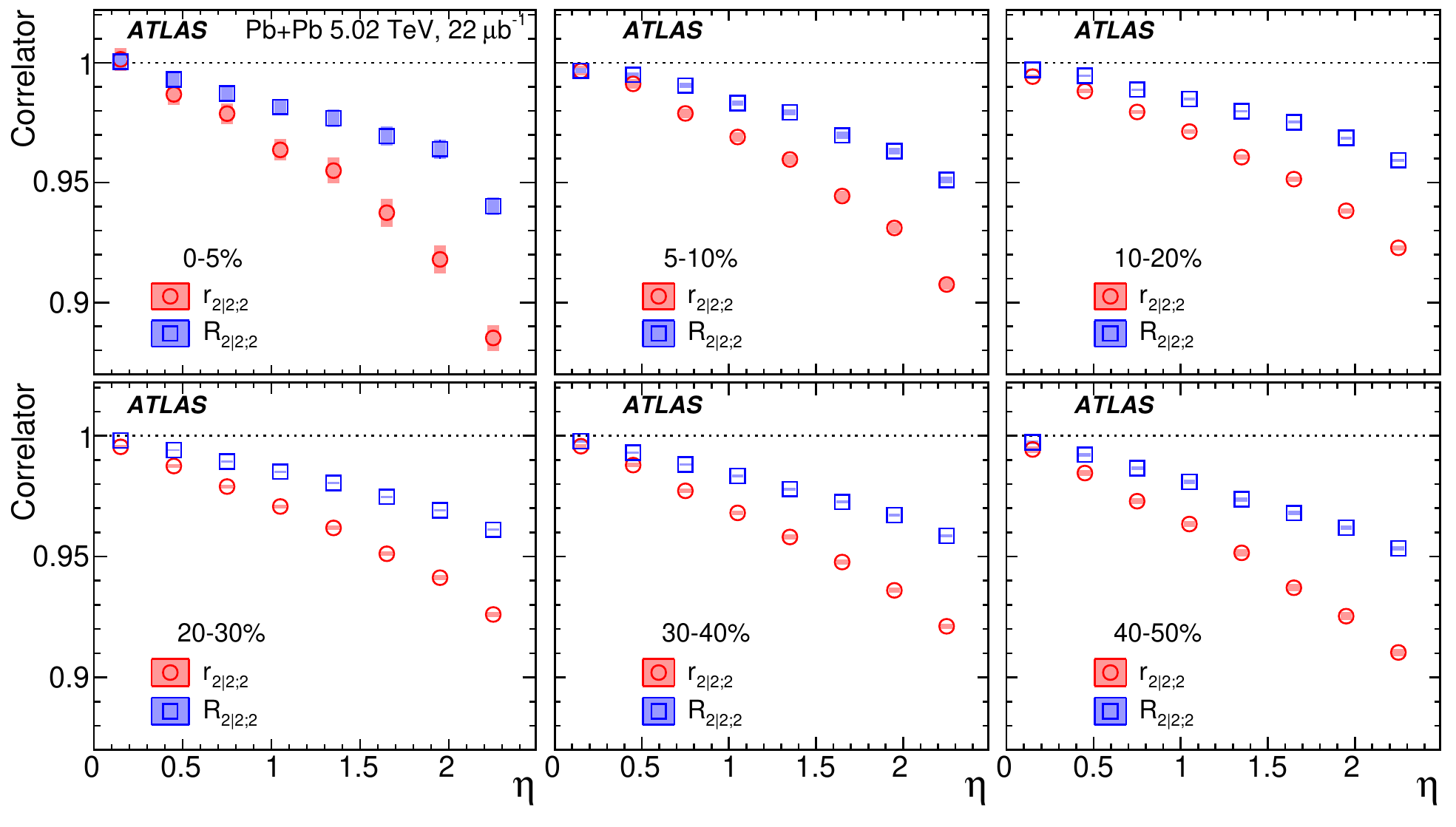}
\end{center}
\caption{\label{fig:14b}  The $r_{2|2;2}(\eta)$ and $R_{2|2;2}(\eta)$ in various centrality intervals for Pb+Pb collisions at 5.02~TeV. The error bars and shaded boxes are statistical and systematic uncertainties, respectively. }
\end{figure}

\begin{figure}[h!]
\begin{center}
\includegraphics[width=0.8\linewidth]{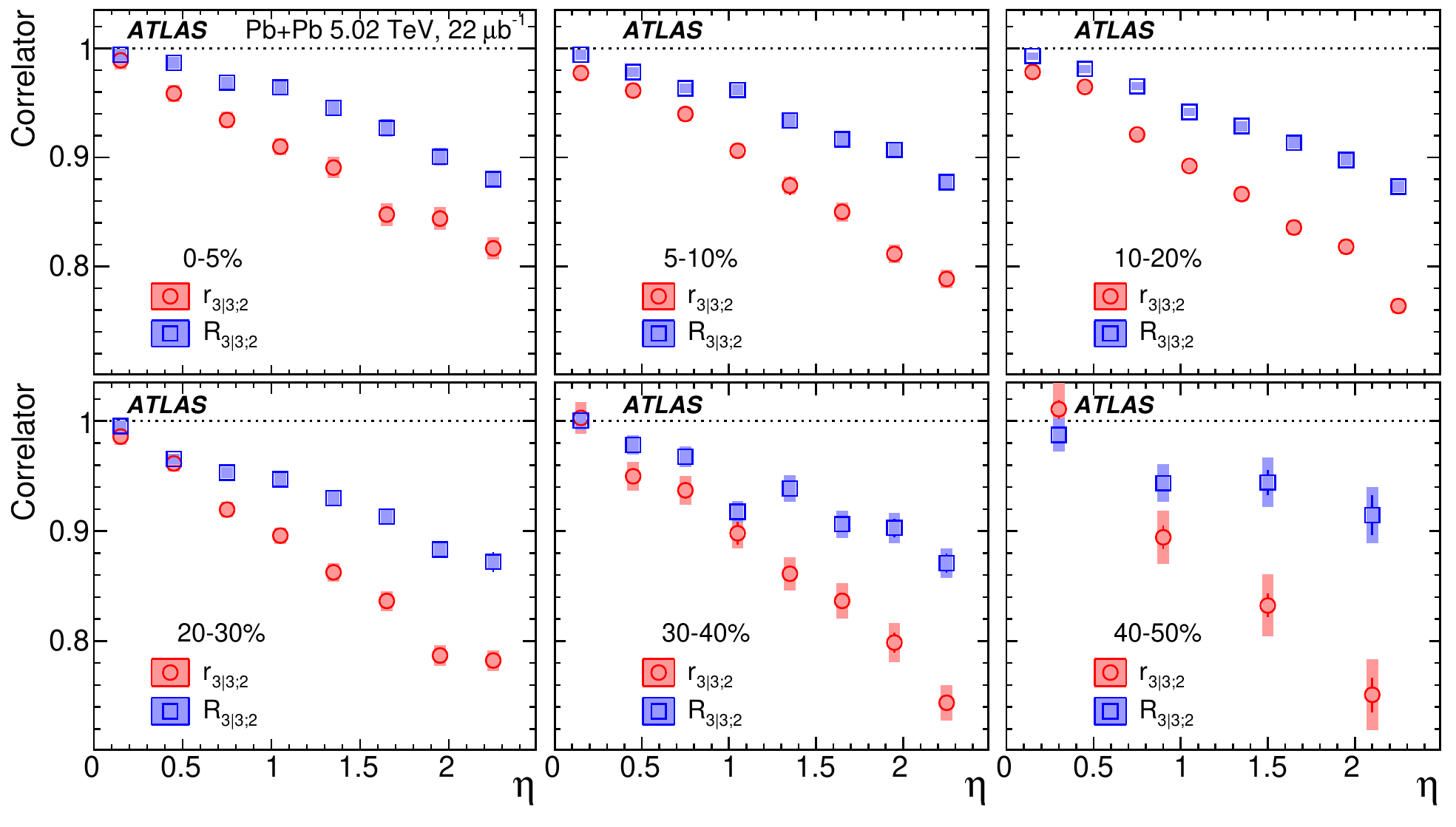}
\end{center}
\caption{\label{fig:14c}  The $r_{3|3;2}(\eta)$ and $R_{3|3;2}(\eta)$ in various centrality intervals for Pb+Pb collisions at 5.02~TeV. The error bars and shaded boxes are statistical and systematic uncertainties, respectively. The data points in 40--50\% centrality interval are rebinned to reduce the uncertainty.}
\end{figure}

Figures~\ref{fig:14b} and \ref{fig:14c} compare the $r_{n|n;2}$ with $R_{n|n;2}$ for $n=2$ and $n=3$, respectively. The decorrelation of $R_{n|n;2}$ is significantly weaker than that for the $r_{n|n;2}$. This is because the $R_{n|n;2}$ is mainly affected by the event-plane twist effects, while the  $r_{n|n;2}$ receives contributions from both FB asymmetry and event-plane twist~\cite{Jia:2017kdq}.

\begin{figure}[h!]
\begin{center}
\includegraphics[width=0.6\linewidth]{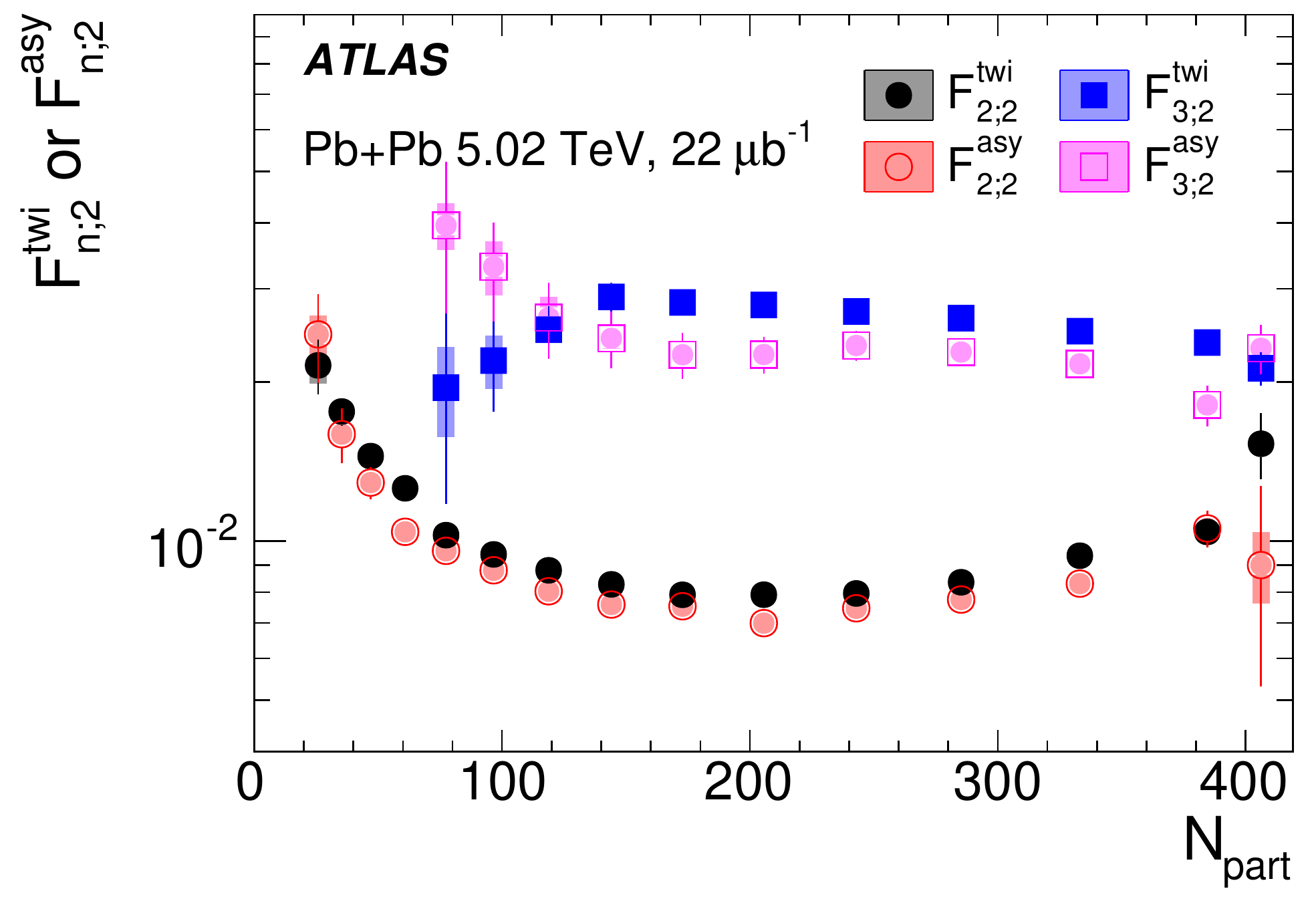}
\end{center}
\caption{\label{fig:15} The estimated event-plane twist component $F_{n;2}^{\textrm{twi}}$ and FB asymmetry component $F_{n;2}^{\textrm{asy}}$ as a function of $\npart$ for $n=2$ and 3 for Pb+Pb collisions at 5.02~TeV. The error bars and shaded boxes are statistical and systematic uncertainties, respectively.}
 \end{figure}

Following the discussion in Section~\ref{sec:obs}, Eqs.~\eqref{eq:3} and \eqref{eq:7}, the measured $F_{n;2}^{\textrm r}$ and $F_{n;2}^{\textrm R}$ values can be used to estimate the separate contributions from FB asymmetry and event-plane twist, $F_{n;2}^{\textrm{asy}}$ and $F_{n;2}^{\textrm{twi}}$, respectively, via the relation:
\begin{eqnarray}
\label{eq:22}
F_{n;2}^{\textrm{twi}} = F_{n;2}^{\textrm R}\;,\;\;\; F_{n;2}^{\textrm{asy}} =  F_{n;2}^{\textrm r}- F_{n;2}^{\textrm R}\;.
\end{eqnarray}
The results are shown in Figure~\ref{fig:15}. The contributions from the two components are similar to each other for $n=2$, for which the harmonic flow arises primarily from the average collision shape, as well as for $n=3$, for which the harmonic flow is driven mainly by fluctuations in the initial geometry.

\subsection{Mixed-harmonics correlation}
\label{sec:res3}
\begin{figure}[h!]
\begin{center}
\includegraphics[width=0.9\linewidth]{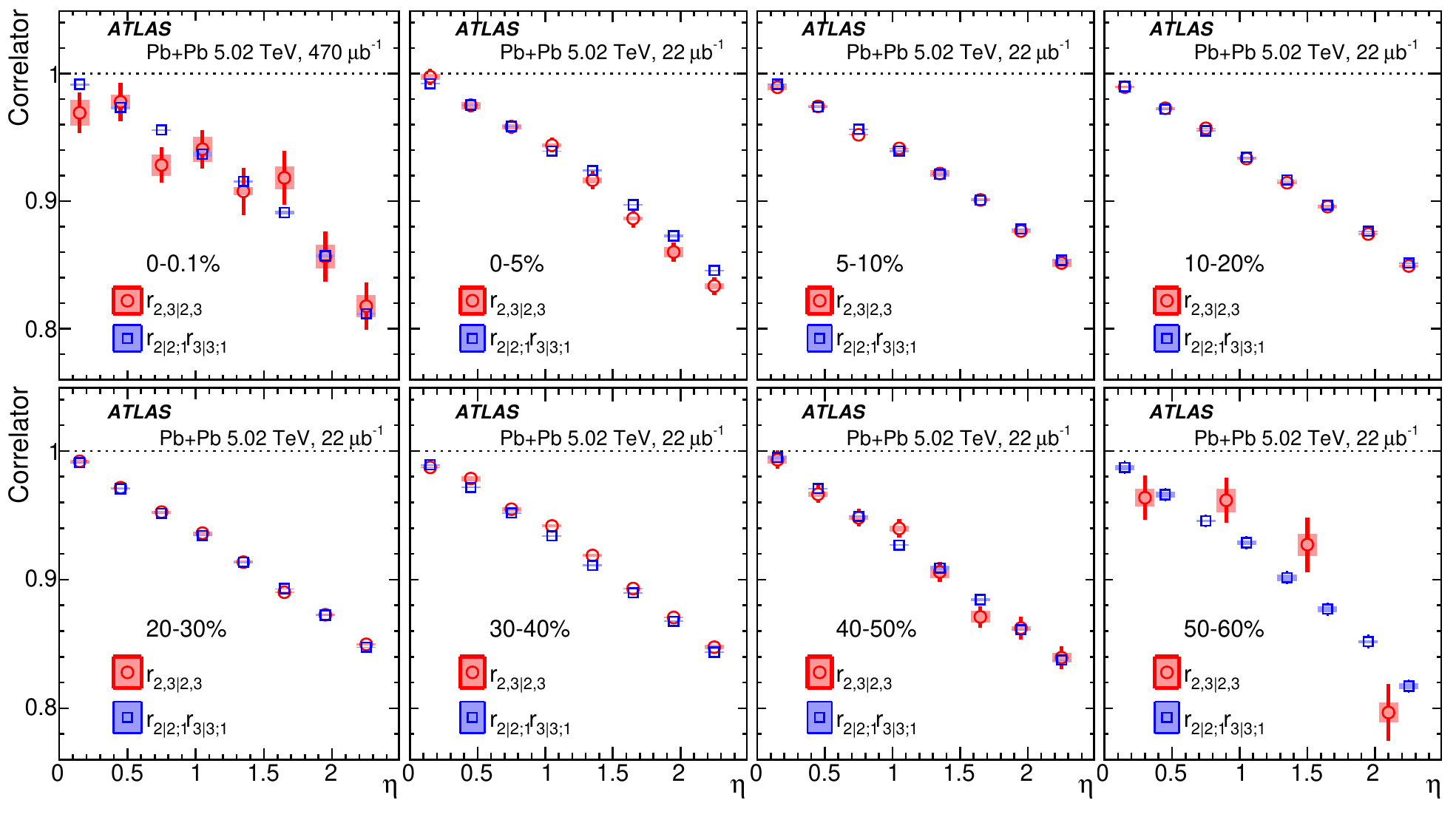}
\end{center}
\caption{\label{fig:17} The $r_{2,3|2,3}$ (circles) and $r_{2|2;1}r_{3|3;1}$ (squares) as a function of $\eta$ for several centrality intervals. The error bars and shaded boxes are statistical and systematic uncertainties, respectively. The $r_{2,3|2,3}$ data in the 50--60\% centrality interval are rebinned to reduce the uncertainty.}
\end{figure}
\begin{figure}[h!]
\begin{center}
\includegraphics[width=0.9\linewidth]{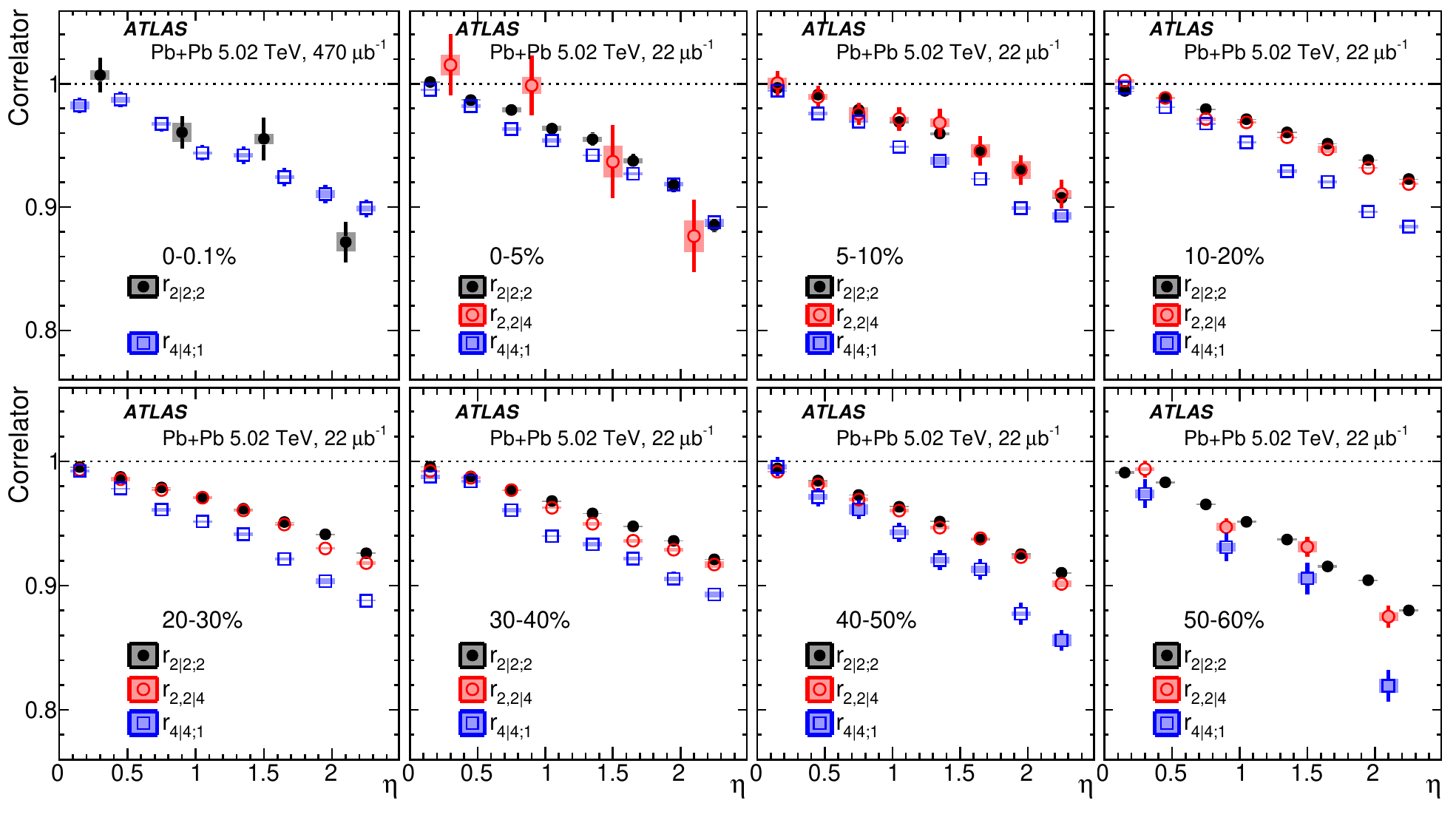}
\end{center}
\caption{\label{fig:18} Comparison of $r_{2|2;2}$, $r_{2,2|4}$ and $r_{4|4;1}$ for several centrality intervals. The error bars and shaded boxes are statistical and systematic uncertainties, respectively. The data points in some centrality intervals are rebinned to reduce the uncertainty.}
\end{figure}
Figure~\ref{fig:17} compares the $r_{2,3|2,3}$ with the product of $r_{2|2;1}$ and $r_{3|3;1}$. The data show that they are consistent with each other, suggesting the previously observed anticorrelation beween $v_2$ and $v_3$ is a property of the entire event~\cite{Aad:2015lwa,Acharya:2017gsw}, and that longitudinal fluctuations of $v_2$ and $v_3$ are uncorrelated. Figure~\ref{fig:18} compares $r_{2|2;2}$ with the mixed-harmonic correlator $r_{2,2|4}$, as well as $r_{4|4;1}$. As discussed in Section~\ref{sec:obs} in the context of the first relation in Eq.~\eqref{eq:11}, if the linear and non-linear components of $v_4$ in Eq.~\eqref{eq:11} are uncorrelated, then $r_{2,2|4}$ would be expected to be similar to $r_{2|2;24}$. This is indeed confirmed by the comparisons of the $\eta$ and centrality dependence of $r_{2|2;2}$ and $r_{2,2|4}$ in Figure~\ref{fig:18}.  Figure~\ref{fig:18} also shows that the $\eta$ dependence for $r_{4|4;1}$ is stronger than for $r_{2|2;2}$ in all centrality intervals, suggesting that the decorrelation effects are stronger for the linear component of $v_4$ than for the nonlinear component (see Eq.~\eqref{eq:13}).

\begin{figure}[h!]
\begin{center}
\includegraphics[width=0.7\linewidth]{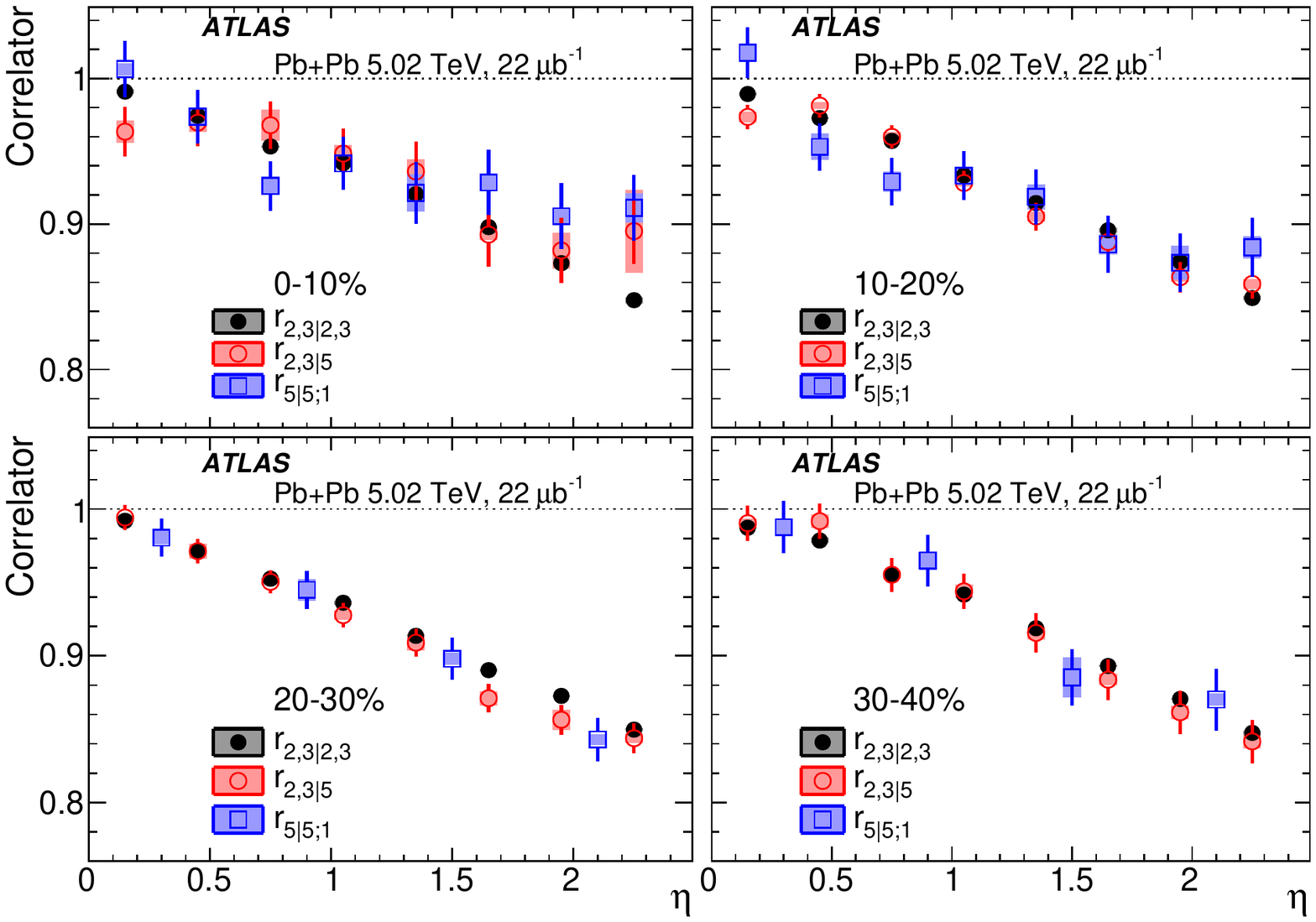}
\end{center}
\caption{\label{fig:19} Comparison of $r_{2,3|2,3}$, $r_{2,3|5}$ and $r_{5|5;1}$ for several centrality intervals. The error bars and shaded boxes are statistical and systematic uncertainties, respectively. The $r_{5|5;1}$ data in some centrality intervals are rebinned to reduce the uncertainty.}
\end{figure}

A similar study of the influence of the linear and nonlinear effects for $v_5$ was also performed, and results are shown in Figure~\ref{fig:19}. The three observables $r_{2,3|2,3}$, $r_{2,3|5}$, and $r_{5|5;1}$ show similar values in all centrality intervals, albeit with large statistical uncertainties. 

\begin{figure}[h!]
\begin{center}
\includegraphics[width=1\linewidth]{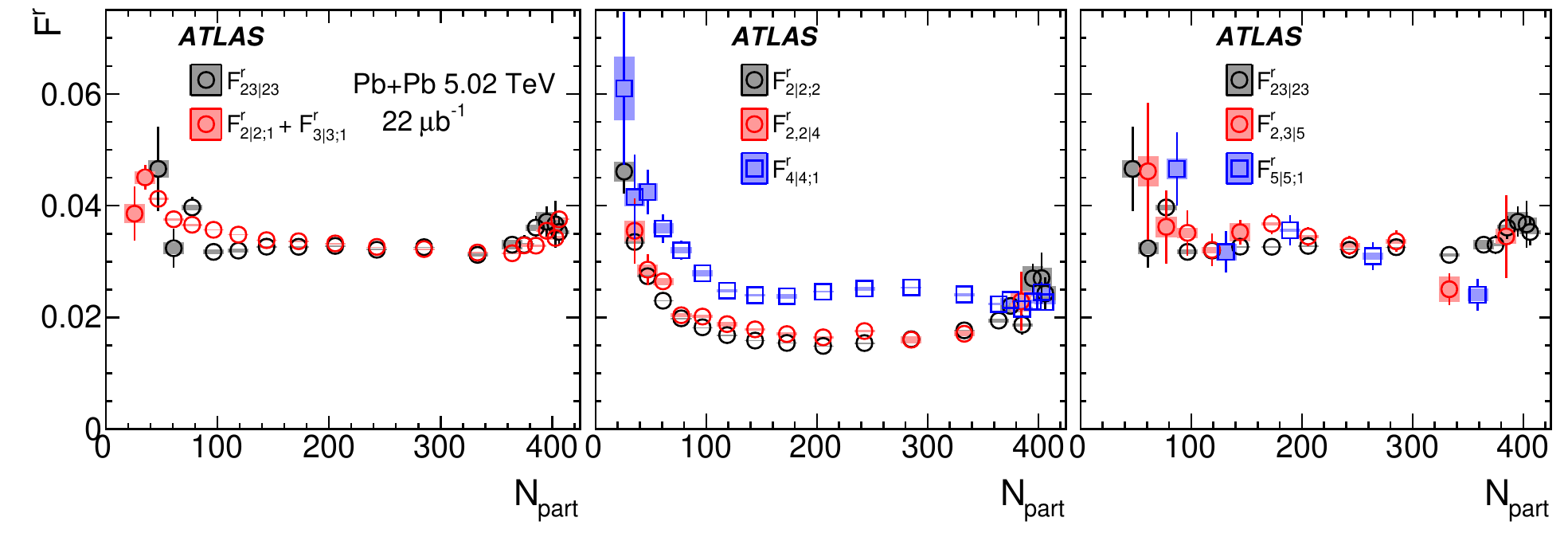}
\end{center}
\caption{\label{fig:20} Comparison of the slopes of the correlators as a function of $\npart$ for three groups of correlators:  $r_{2,3|2,3}$ and $r_{2|2;1}r_{3|3;1}$ (for which the slope is $F_{2|2;1}+F_{3|3;1}$) in Figure~\ref{fig:17} (left panel), $r_{2|2;2}$, $r_{2,2|4}$ and $r_{4|4;1}$ in Figure~\ref{fig:18} (middle panel), and  $r_{2,3|2,3}$, $r_{2,3|5}$ and $r_{5|5;1}$ in Figure~\ref{fig:19} (right panel). The error bars and shaded boxes are statistical and systematic uncertainties, respectively.}
\end{figure}

The decorrelations shown in Figures~\ref{fig:17}--\ref{fig:19} can be quantified by calculating the slopes of the distributions in each centrality interval and presenting the results as a function of centrality. Following the example for $r_{n|n;k}$, the slopes for the mixed-harmonic correlators are obtained via the linear regression procedure of Eqs.~\eqref{eq:20} and \eqref{eq:20b}:
\begin{eqnarray}
\label{eq:23}
r_{2,3|2,3} = 1-2F_{2,3|2,3}^{\textrm r}\;\eta\;,\;\;r_{2,2|4} = 1-2F_{2,2|4}^{\textrm r}\;\eta\;,\;\;r_{2,3|5} = 1-2F_{2,3|5}^{\textrm r}\;\eta\;.
\end{eqnarray}
The results are summarised in Figure~\ref{fig:20}, with each panel corresponding to the slopes of distributions in Figures~\ref{fig:17}, \ref{fig:18}, and \ref{fig:19}, respectively. The only significant difference is seen between $F_{4|4;1}$ and $F_{2|2;2}$ or $F_{2,2|4}$.

\pagebreak
\section{Summary}
Measurements of longitudinal flow correlations for charged particles are presented in the pseudorapidity range $|\eta|<2.4$ using 7~$\mu$b$^{-1}$ and 470~$\mu$b$^{-1}$ of Pb+Pb data at $\sqrtsnn=2.76$ and 5.02~TeV, respectively, recorded by the ATLAS detector at the LHC. The factorisation of two-particle azimuthal correlations into single-particle flow harmonics $v_n$ is found to be broken, and the amount of factorisation breakdown increases approximately linearly as a function of the $\eta$ separation between the two particles. The slope of this dependence is nearly independent of centrality and $\pT$ for $n>2$. However, for $n=2$ the effect is smallest in mid-central collisions and increases toward more central or more peripheral collisions, and in central collisions the effect also depends strongly on $\pT$. Furthermore, the effect is found to be larger at 2.76~TeV than 5.02~TeV for all harmonics, which cannot be explained entirely by the change in the beam rapidity. 

The higher moments of the $\eta$-dependent flow correlations are also measured and the corresponding linear coefficients of the $\eta$ dependence are extracted. The coefficient for the $k^{\textrm{th}}$-moment of $v_n$ scales with $k$ for $n>2$, but scales faster than $k$ for $n=2$. The factorisation breakdown is separated into contributions from forward-backward asymmetry of the flow magnitude and event-plane twist, which are found to be comparable to each other. 

The longitudinal flow correlations are also measured between harmonic flows of different order. The correlation of $v_2v_3$ between two $\eta$ ranges is found to factorise into the product of the correlation for $v_2$ and the correlation for $v_3$, suggesting that the longitudinal fluctuations of $v_2$ and $v_3$ are independent of each other. The correlations between $v_4$ and $v_2^2$ suggest that the longitudinal fluctuations of $v_4$ have a significant nonlinear contribution from $v_2$, i.e.\ $v_4\propto v_2^2$. Similarly, the correlations between $v_5$ and $v_2v_3$ suggest that the longitudinal fluctuations of $v_5$ are driven by the nonlinear contribution from $v_2v_3$, i.e.\ $v_5\propto v_2v_3$. The results presented in this paper provide new insights into the fluctuations and correlations of harmonic flow in the longitudinal direction, which can be used to improve full three-dimensional viscous hydrodynamic models.

\section*{Acknowledgements}

We thank CERN for the very successful operation of the LHC, as well as the
support staff from our institutions without whom ATLAS could not be
operated efficiently.

We acknowledge the support of ANPCyT, Argentina; YerPhI, Armenia; ARC, Australia; BMWFW and FWF, Austria; ANAS, Azerbaijan; SSTC, Belarus; CNPq and FAPESP, Brazil; NSERC, NRC and CFI, Canada; CERN; CONICYT, Chile; CAS, MOST and NSFC, China; COLCIENCIAS, Colombia; MSMT CR, MPO CR and VSC CR, Czech Republic; DNRF and DNSRC, Denmark; IN2P3-CNRS, CEA-DRF/IRFU, France; SRNSF, Georgia; BMBF, HGF, and MPG, Germany; GSRT, Greece; RGC, Hong Kong SAR, China; ISF, I-CORE and Benoziyo Center, Israel; INFN, Italy; MEXT and JSPS, Japan; CNRST, Morocco; NWO, Netherlands; RCN, Norway; MNiSW and NCN, Poland; FCT, Portugal; MNE/IFA, Romania; MES of Russia and NRC KI, Russian Federation; JINR; MESTD, Serbia; MSSR, Slovakia; ARRS and MIZ\v{S}, Slovenia; DST/NRF, South Africa; MINECO, Spain; SRC and Wallenberg Foundation, Sweden; SERI, SNSF and Cantons of Bern and Geneva, Switzerland; MOST, Taiwan; TAEK, Turkey; STFC, United Kingdom; DOE and NSF, United States of America. In addition, individual groups and members have received support from BCKDF, the Canada Council, CANARIE, CRC, Compute Canada, FQRNT, and the Ontario Innovation Trust, Canada; EPLANET, ERC, ERDF, FP7, Horizon 2020 and Marie Sk{\l}odowska-Curie Actions, European Union; Investissements d'Avenir Labex and Idex, ANR, R{\'e}gion Auvergne and Fondation Partager le Savoir, France; DFG and AvH Foundation, Germany; Herakleitos, Thales and Aristeia programmes co-financed by EU-ESF and the Greek NSRF; BSF, GIF and Minerva, Israel; BRF, Norway; CERCA Programme Generalitat de Catalunya, Generalitat Valenciana, Spain; the Royal Society and Leverhulme Trust, United Kingdom.

The crucial computing support from all WLCG partners is acknowledged gratefully, in particular from CERN, the ATLAS Tier-1 facilities at TRIUMF (Canada), NDGF (Denmark, Norway, Sweden), CC-IN2P3 (France), KIT/GridKA (Germany), INFN-CNAF (Italy), NL-T1 (Netherlands), PIC (Spain), ASGC (Taiwan), RAL (UK) and BNL (USA), the Tier-2 facilities worldwide and large non-WLCG resource providers. Major contributors of computing resources are listed in Ref.~\cite{ATL-GEN-PUB-2016-002}.

\appendix
\part*{Appendix}
\begin{figure}[h!]
\begin{center}
\includegraphics[width=0.9\linewidth]{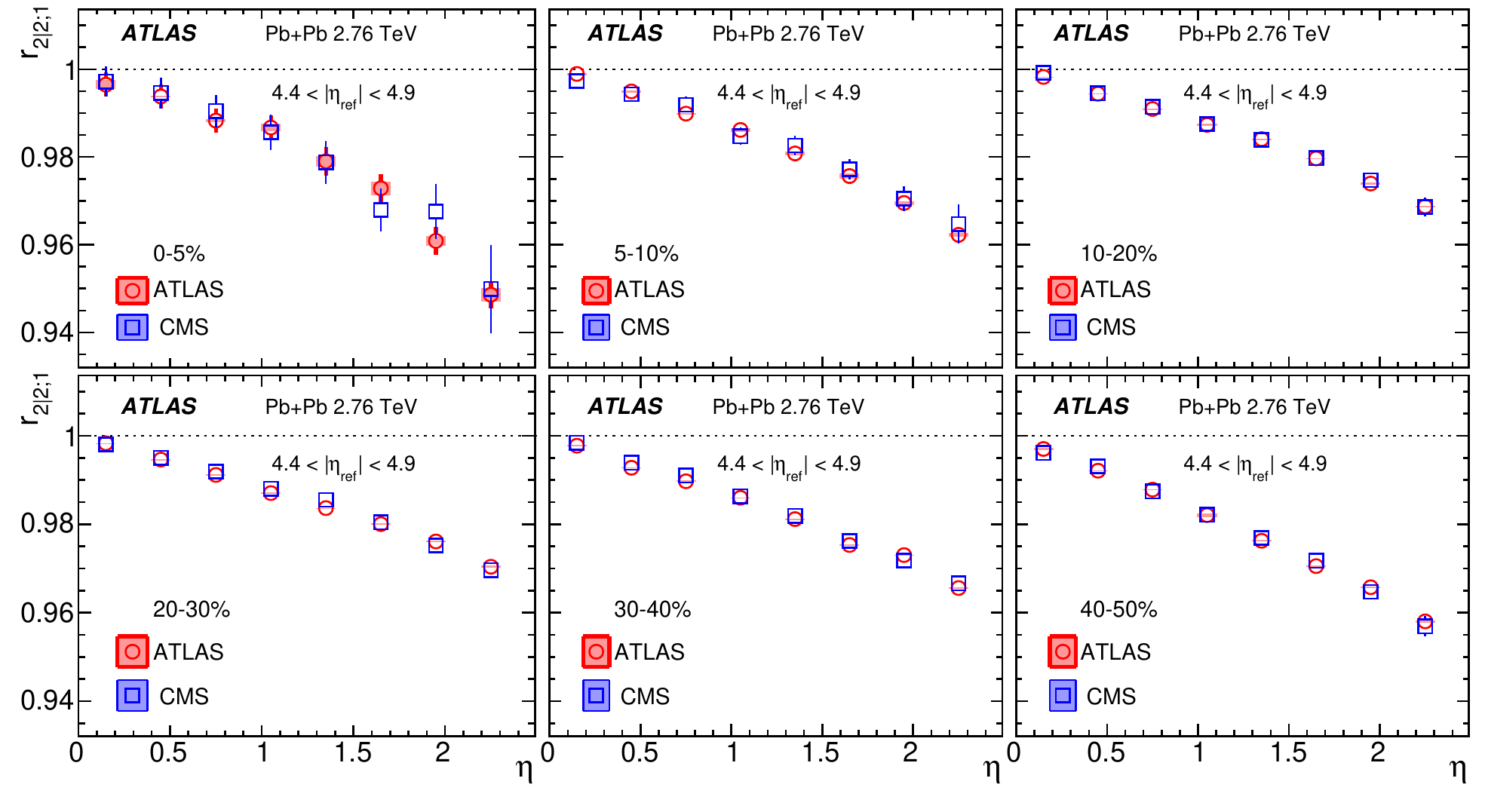}
\end{center}
\caption{\label{fig:1a} The values of $r_{2|2;1}$ measured by ATLAS and by CMS~\cite{Khachatryan:2015oea} for Pb+Pb collisions at 2.76~TeV, for the same reference pseudorapidity $4.4< |\etaref|<4.9$.}
\end{figure}

\begin{figure}[h!]
\begin{center}
\includegraphics[width=0.9\linewidth]{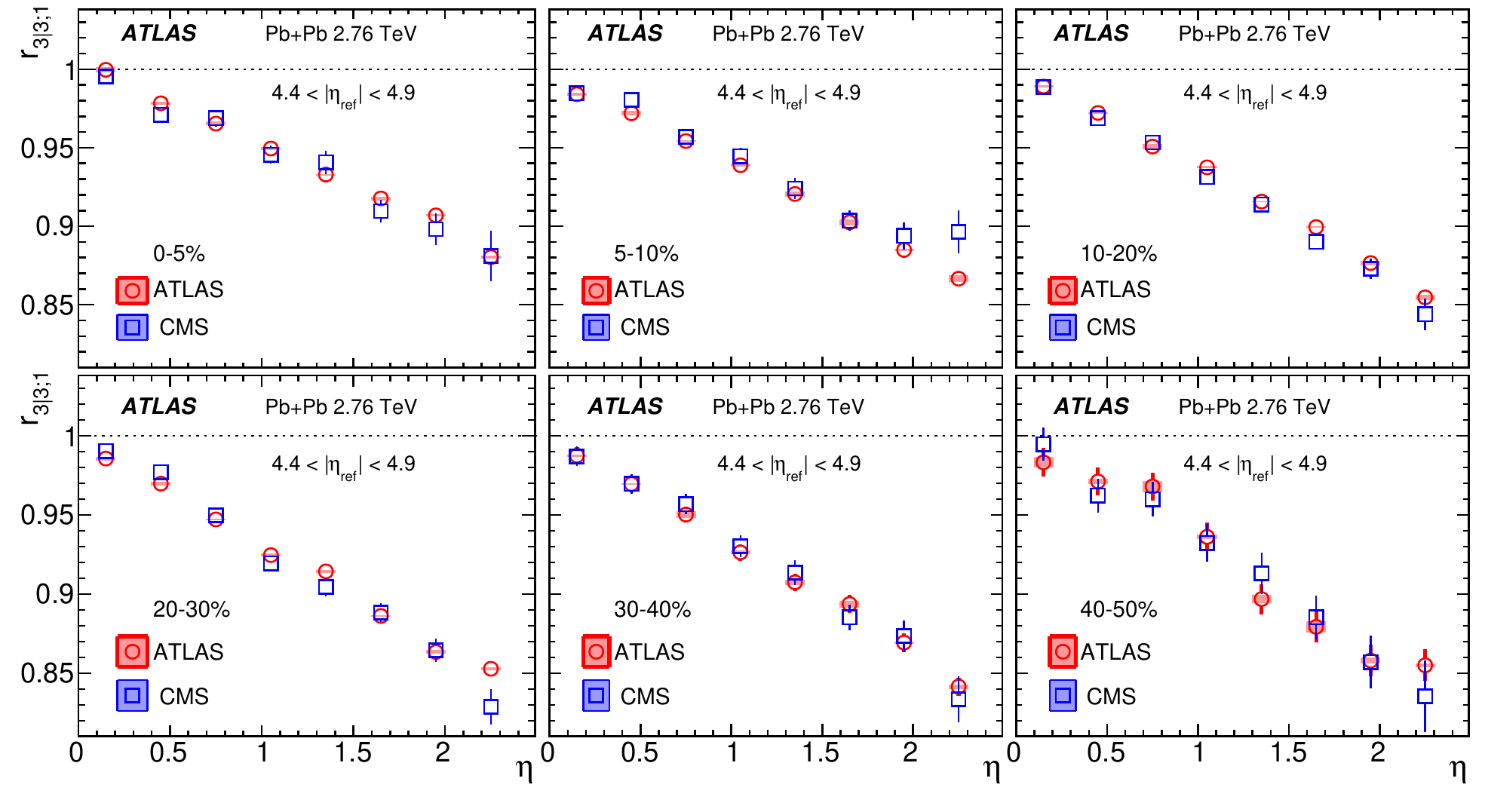}
\end{center}
\caption{\label{fig:2a} The values of $r_{3|3;1}$ measured by ATLAS and by CMS~\cite{Khachatryan:2015oea} for Pb+Pb collisions at 2.76~TeV, for the same reference pseudorapidity $4.4< |\etaref|<4.9$.}
\end{figure}

\begin{figure}[h!]
\begin{center}
\includegraphics[width=0.9\linewidth]{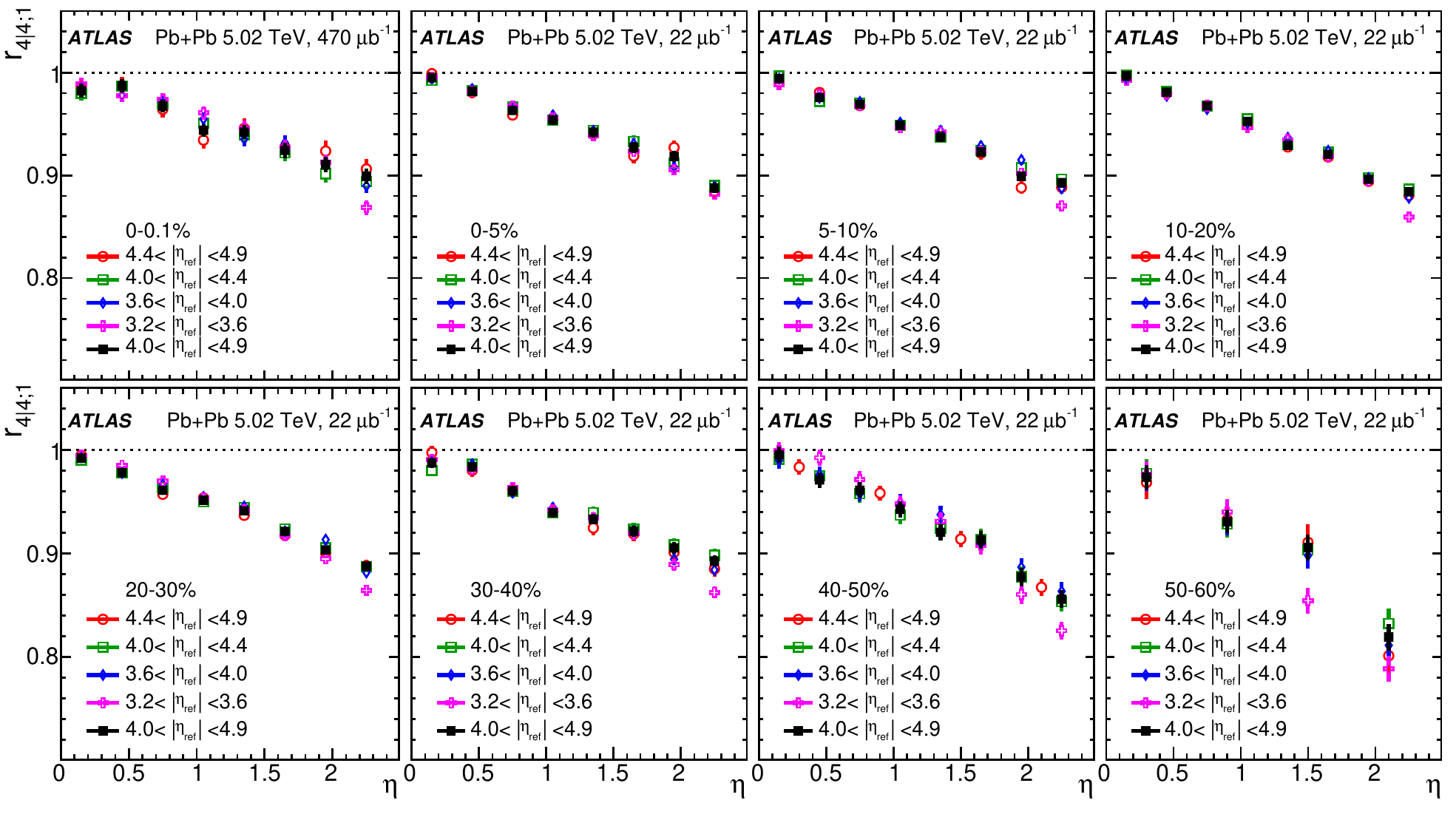}
\end{center}\vspace*{-0.5cm}
\caption{\label{fig:3a}  The $r_{4|4;1}(\eta)$ measured for several $\etaref$ ranges for Pb+Pb collisions at 5.02~TeV. Each panel represents one centrality range. The error bars are statistical only. The data points in some centrality intervals are rebinned to reduce the uncertainty.}
\end{figure}

\begin{figure}[h!]
\begin{center}
\includegraphics[width=0.9\linewidth]{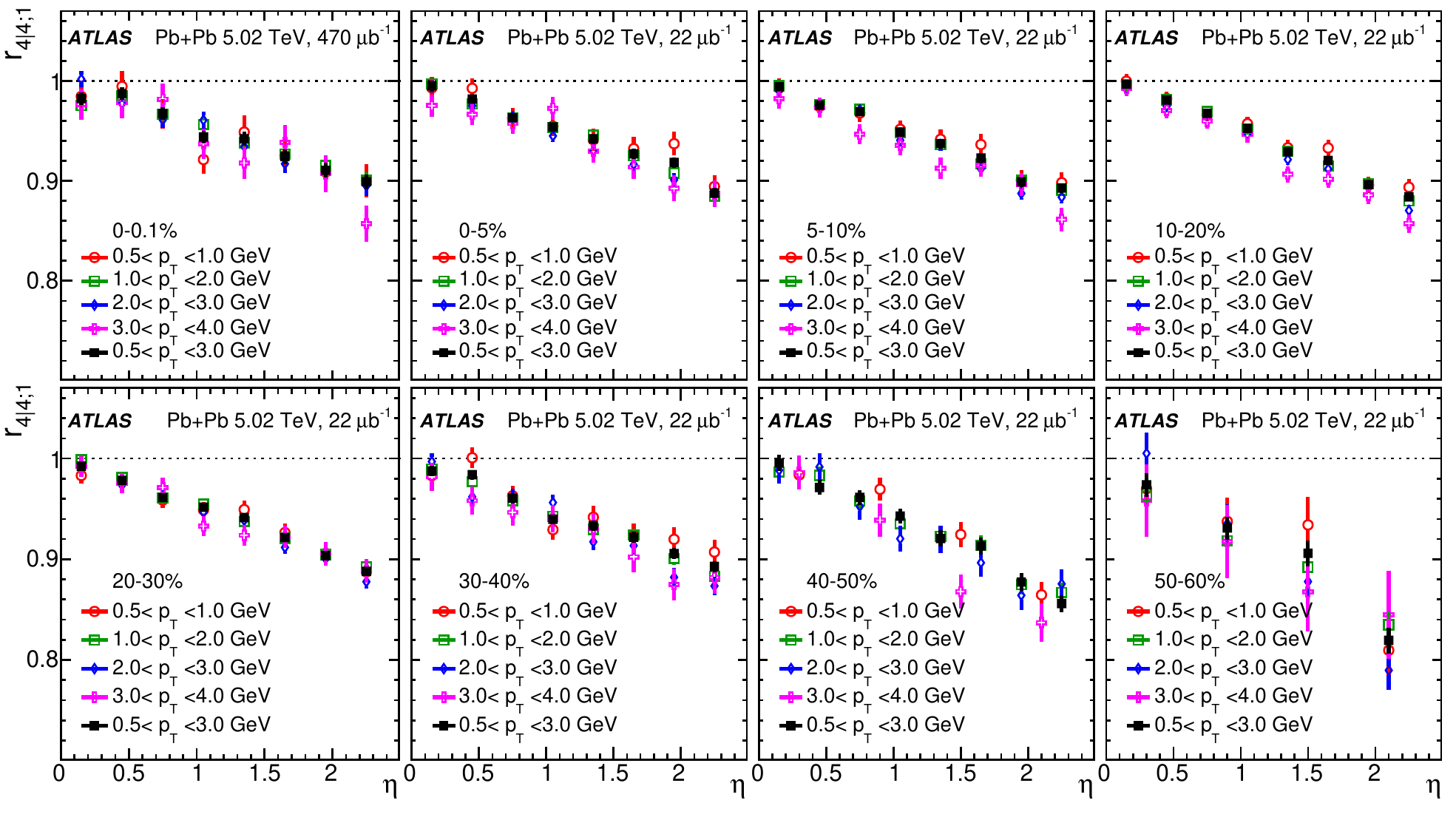}
\end{center}\vspace*{-0.5cm}
\caption{\label{fig:4a} The $r_{4|4;1}(\eta)$ measured in several $\pT$ ranges for Pb+Pb collisions at 5.02~TeV. Each panel shows the results for one centrality range.  The error bars are statistical only. The data points in some centrality intervals are rebinned to reduce the uncertainty.}
\end{figure}

\begin{figure}[h!]
\begin{center}
\includegraphics[width=0.8\linewidth]{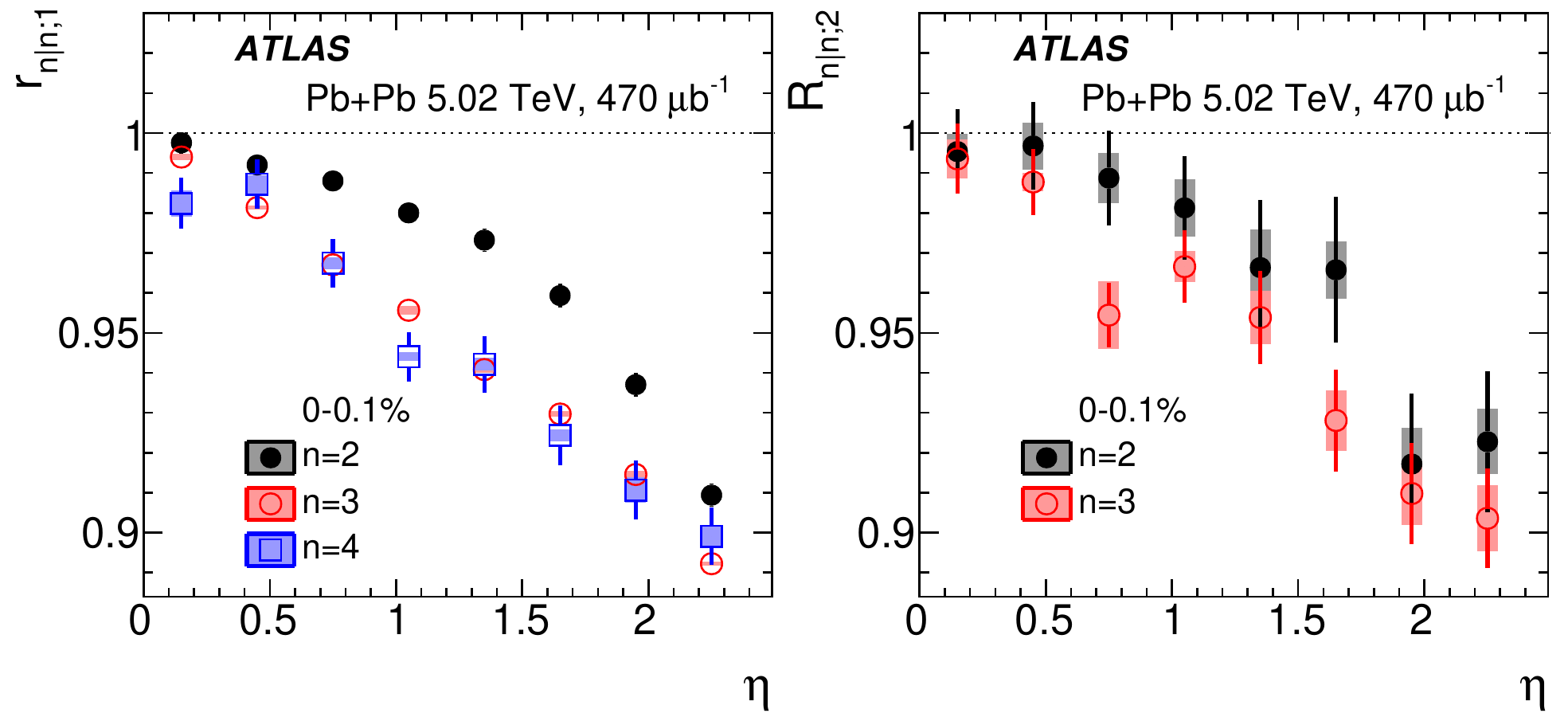}
\end{center}
\caption{\label{fig:5a}  The $r_{n|n;1}(\eta)$ (left panel) and $R_{n|n;2}(\eta)$ (right panel) in ultra-central Pb+Pb collisions at 5.02~TeV. The error bars and shaded boxes are statistical and systematic uncertainties, respectively.}
\end{figure}

\addcontentsline{toc}{part}{Appendix}
Figure~\ref{fig:1a} and \ref{fig:2a} show a comparison of $r_{2|2;1}$ and $r_{3|3;1}$ between ATLAS and CMS for Pb+Pb collisions at 2.76 TeV, where the ATLAS $\etaref$ is chosen to be $4.4<|\etaref|<4.9$ to match that of the CMS Collaboration. Excellent agreement is observed. Figure~\ref{fig:3a} and \ref{fig:4a} show the detailed $\pT$ and $\etaref$ dependence of $r_{4|4;1}$; these figures complement Figures~\ref{fig:1}--\ref{fig:4}. Figure~\ref{fig:5a} compiles the results of $r_{n|n;1}$ and $R_{n|n;2}$ for 0--0.1\% ultra-central collisions. 


\clearpage


\bibliographystyle{atlasBibStyleWithTitle}
\bibliography{hion-2016-04}



\begin{flushleft}
{\Large The ATLAS Collaboration}

\bigskip

M.~Aaboud$^\textrm{\scriptsize 137d}$,
G.~Aad$^\textrm{\scriptsize 88}$,
B.~Abbott$^\textrm{\scriptsize 115}$,
O.~Abdinov$^\textrm{\scriptsize 12}$$^{,*}$,
B.~Abeloos$^\textrm{\scriptsize 119}$,
S.H.~Abidi$^\textrm{\scriptsize 161}$,
O.S.~AbouZeid$^\textrm{\scriptsize 139}$,
N.L.~Abraham$^\textrm{\scriptsize 151}$,
H.~Abramowicz$^\textrm{\scriptsize 155}$,
H.~Abreu$^\textrm{\scriptsize 154}$
% ...
\bigskip
\\
% ...
$^{12}$ Institute of Physics, Azerbaijan Academy of Sciences, Baku, Azerbaijan\\
% ...
$^{88}$ CPPM, Aix-Marseille Universit{\'e} and CNRS/IN2P3, Marseille, France\\
% ...
$^{115}$ Homer L. Dodge Department of Physics and Astronomy, University of Oklahoma, Norman OK, United States of America\\
% ...
$^{119}$ LAL, Univ. Paris-Sud, CNRS/IN2P3, Universit{\'e} Paris-Saclay, Orsay, France\\
% ...
$^{137}$ $^{(a)}$ Facult{\'e} des Sciences Ain Chock, R{\'e}seau Universitaire de Physique des Hautes Energies - Universit{\'e} Hassan II, Casablanca; $^{(b)}$ Centre National de l'Energie des Sciences Techniques Nucleaires, Rabat; $^{(c)}$ Facult{\'e} des Sciences Semlalia, Universit{\'e} Cadi Ayyad, LPHEA-Marrakech; $^{(d)}$ Facult{\'e} des Sciences, Universit{\'e} Mohamed Premier and LPTPM, Oujda; $^{(e)}$ Facult{\'e} des sciences, Universit{\'e} Mohammed V, Rabat, Morocco\\
% ...
$^{139}$ Santa Cruz Institute for Particle Physics, University of California Santa Cruz, Santa Cruz CA, United States of America\\
% ...
$^{151}$ Department of Physics and Astronomy, University of Sussex, Brighton, United Kingdom\\
% ...
$^{154}$ Department of Physics, Technion: Israel Institute of Technology, Haifa, Israel\\
$^{155}$ Raymond and Beverly Sackler School of Physics and Astronomy, Tel Aviv University, Tel Aviv, Israel\\
% ...
$^{161}$ Department of Physics, University of Toronto, Toronto ON, Canada\\
% ...
$^{a}$ Also at Department of Physics, King's College London, London, United Kingdom\\
$^{b}$ Also at Institute of Physics, Azerbaijan Academy of Sciences, Baku, Azerbaijan\\
$^{c}$ Also at Novosibirsk State University, Novosibirsk, Russia\\
$^{d}$ Also at TRIUMF, Vancouver BC, Canada\\
$^{e}$ Also at Department of Physics {\&} Astronomy, University of Louisville, Louisville, KY, United States of America\\
% ...
$^{*}$ Deceased
\end{flushleft}


\end{document}